# HERMES: towards an integrated toolbox to characterize functional and effective brain connectivity

Guiomar Niso[1,*], Ricardo Bruña[1], Ernesto Pereda[2],

Ricardo Gutiérrez[1], Ricardo Bajo[1], Fernando Maestú[1], Francisco del-Pozo[1]

[1] Centre for Biomedical Technology, Technical University of Madrid, Spain

[2] Electrical Engineering and Bioengineering Group, Dept. of Basic Physics, University of La Laguna, Tenerife, Spain

[*] Corresponding author. Address for correspondence: guiomar.niso@ctb.upm.es

## ABSTRACT

The analysis of the interdependence between time series has become an important field of research in the last years, mainly as a result of advances in the characterization of dynamical systems from the signals they produce, the introduction of concepts such as generalized and phase synchronization and the application of information theory to time series analysis. In neurophysiology, different analytical tools stemming from these concepts have added to the 'traditional' set of linear methods, which includes the cross-correlation and the coherency function in the time and frequency domain, respectively, or more elaborated tools such as Granger Causality.

This increase in the number of approaches to tackle the existence of functional (FC) or effective connectivity (EC) between two (or among many) neural networks, along with the mathematical complexity of the corresponding time series analysis tools, makes it desirable to arrange them into a unified, easy-to-use software package. The goal is to allow neuroscientists, neurophysiologists and researchers from related fields to easily access and make use of these analysis methods from a single integrated toolbox.

Here we present HERMES (http://hermes.ctb.upm.es), a toolbox for the Matlab® environment (The Mathworks, Inc), which is designed to study functional and effective brain connectivity from neurophysiological data such as multivariate EEG and/or MEG records. It includes also visualization tools and statistical methods to address the problem of multiple comparisons. We believe that this toolbox will be very helpful to all the researchers working in the emerging field of brain connectivity analysis.

# CONTENT INDEX

# FIGURES

# TABLES

# HERMES: towards an integrated toolbox to characterize functional and effective brain connectivity

Guiomar Niso[1,*], Ricardo Bruña[1], Ernesto Pereda[2],

Ricardo Gutiérrez[1], Ricardo Bajo[1], Fernando Maestú[1], Francisco del-Pozo[1]

[1] Centre for Biomedical Technology, Technical University of Madrid, Spain

[2] Electrical Engineering and Bioengineering Group, Dept. of Basic Physics, University of La Laguna, Tenerife, Spain

[*] Corresponding author. Address for correspondence: guiomar.niso@ctb.upm.es

## 1. INTRODUCTION

The analysis of the interdependence between time series has become an important field of research, partly as a result of advances in the characterization of dynamical systems from the signals they produce, and the introduction of concepts such as generalized (GS) and phase synchronization (PS). In neurophysiology, different analytical tools stemming from these and related concepts (Pereda et al. 2005) have added to the "traditional" set of linear methods of multivariate time series analysis, such as the cross-correlation or the coherence. The popularity of these tools has grown in parallel with that of the idea of connectivity as one of the crucial aspects underlying information processing in the brain. Brain connectivity is an elusive concept that refers to different interrelated aspects of brain organization (see, e.g., (Horwitz, 2003) for a critical review), and is normally divided into three different categories: anatomical or structural, functional (FC) and effective connectivity (EC). Anatomical connectivity refers to a network of *physical connections* linking sets of neurons or neuronal elements, and has to do with the anatomical structure of brain networks. However, FC refers to the statistical dependence between the signals stemming from two (or among many) distinct units within a nervous system (from single neurons to whole neural networks), while EC refers to the causal interactions between (or among) them (Friston 1994; 2011). Note that here FC does not entail the existence of any physical connection between these networks (i.e., in terms of tracts or fibres linking the two brain sites). It only refers to the existence of a relationship between the corresponding signals. In turn, the causal relationship that defines EC is also reflected in the signals as the existence, e.g., of a coherent time lag between them or an asymmetry in the dependence between their reconstructed state spaces. That is why FC/EC can be tackled from multivariate neurophysiological signals with the help of tools for the analysis of the interdependence between time series. Roughly speaking, FC is assessed by those (symmetric) tools that measures the existence of any type of covariance (whether linear or nonlinear) between two neurophysiological signals without providing any causal information (good examples are the traditional linear methods mentioned above), whereas for the assessment of EC one needs time series techniques that do provide causal information, such as Granger causality (Granger, 1969) or transfer entropy (Schreiber, 2000).

Recently, there has been an outburst of toolboxes that include indexes of brain connectivity, toolboxes that are made publicly available and published in the literature (Oostenveld et al., 2011; Rose, Otto, & Dittrich, 2008; Seth, 2010; Tadel et al., 2011; Zhou et al., 2009). However, most of them either focus on a special type of connectivity indexes (e.g., linear indexes (Seth, 2010)) and/or include only a subset of indexes as part of a more general purpose toolbox whose main aim is, say, the analysis of EEG and/or MEG (Delorme & Makeig 2004; Delorme et al. 2011; Oostenveld et al. 2011; Tadel et al. 2011). Yet we feel that the increase in the number of time series analysis tools to study FC / EC in the brain, along with their mathematical complexity, makes it desirable to arrange them into a (still missing) single, unified toolbox that allow neuroscientists, neurophysiologists and researchers from related fields to easily access and make use of them. Consequently, we hereby present a new toolbox called HERMES, running under the crossplatform Matlab® environment, which encompasses several of the most common indexes for the assessment of FC and EC. Besides, the toolbox also includes visualization routines and two different advanced statistical methods that address the problem of multiple comparisons, which are also very useful tools for the analysis of connectivity in multivariate neuroimage data sets.

HERMES is the Spanish abbreviation for *"HERramientas de MEdidas de Sincronización"*, which roughly translates to English as *"Tools for the Assessment of Synchronization"*. But HERMES is also the name of the messenger of the gods in Greek mythology, the guide to the Underworld (the study of theory and practice of interpretation –Hermeneutics- is also named after him). By naming the toolbox after such deity, we want to highlight the purpose that inspired it: to allow researchers not familiar with the underlying mathematics gaining access to different connectivity measures and analysis tools.

## 2. PROJECT CREATION

As stated above, HERMES is a Matlab® toolbox. Thus, it has to be launched from the Matlab® environment. The simplest and most straightforward way of using HERMES is through its graphical user interface (GUI, see Fig. 1), which is invoked by typing, in the command line:

» **HERMES**

This opens the GUI and allows the user to start creating a new project.

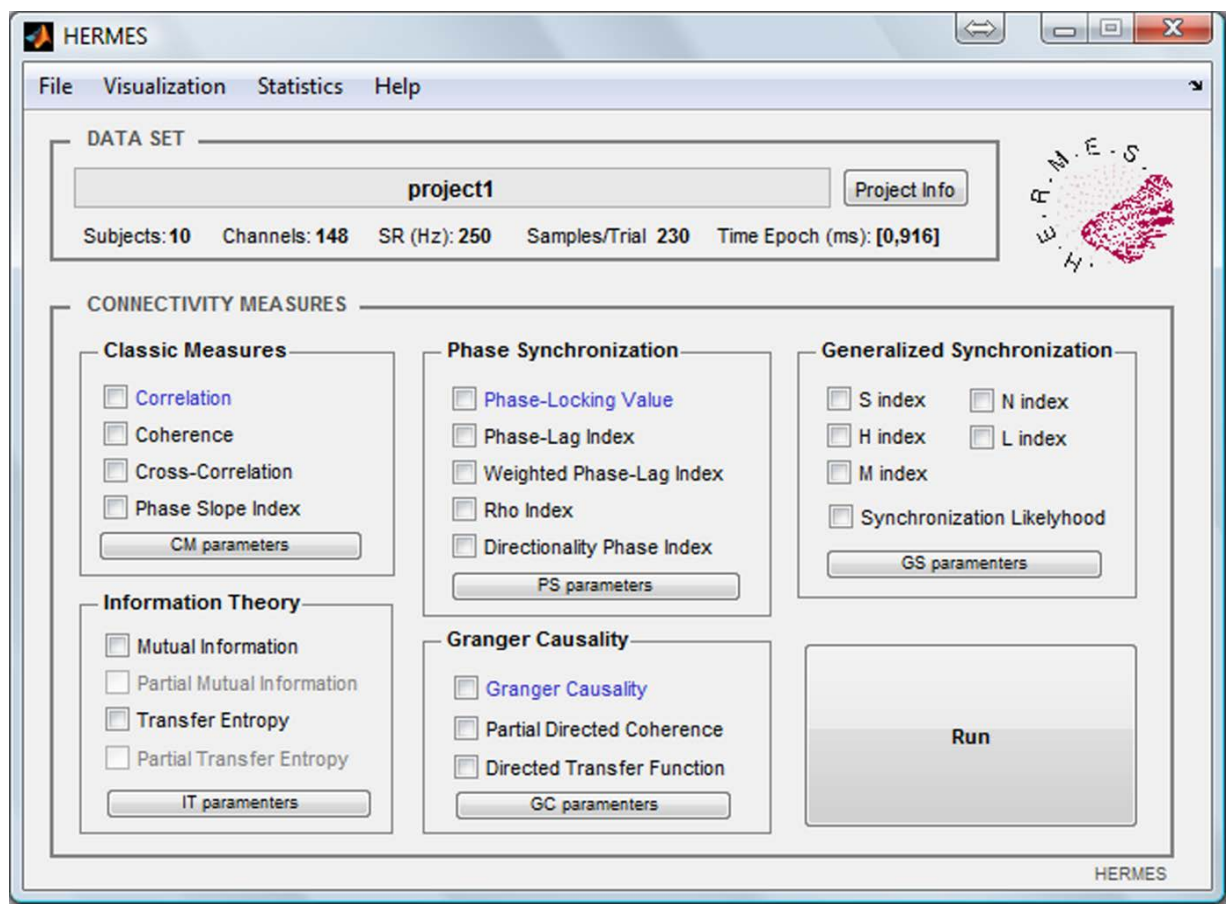

**Fig. 1 HERMES graphical user interface**

HERMES Graphical User Interface (GUI) is divided in the 'data set' zone (top), which contains the project's most relevant data, and the 'connectivity measures' zone (bottom), which gathers the different types of indexes the toolbox can compute

HERMES is a project-based toolbox. This means that a new project must be created before we can start working on the data. The project contains the data matrices to be analysed, the metadata of these matrices (i.e. sampling rate, pre-stimulus time, conditions, etc.) and all (if any) previously calculated indexes.

As commented in the Introduction, the main purpose of HERMES is the analysis of brain FC and EC. Therefore, it does not include any artefact-removal, detrending or any similar pre-processing tools, which are already available in popular Matlab® toolboxes oriented to specific neuroimaging data analysis (Delorme et al. 2011; Tadel et al. 2011). Thus, loaded data should be clean (i.e. artefact-free) and, if necessary, epoched. For those indexes that require data filtering (i.e. PS indexes), the signal will be internally filtered using a finite impulse response filter of optimal order (i.e., one third of the window size, in samples).

A project may contain a single data matrix or data obtained from different subjects and/or under different conditions. HERMES can load both matrices and *FieldTrip* structures (Oostenveld et al., 2011) stored in MAT files. If more than one data file is loaded, HERMES will ask the user for information about the subject(s), group(s) and condition(s) of each file, by means of the *Data labelling* panel (see Fig. 2).

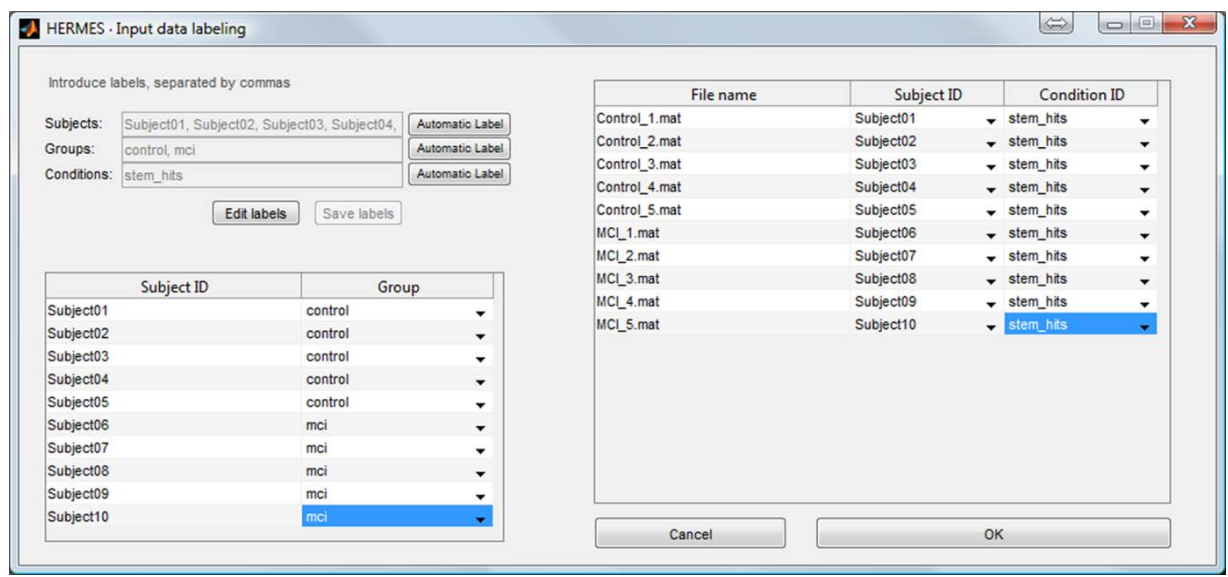

**Fig. 2 Data labelling panel**

Data labelling panel to classify your project's subjects. They can belong to different groups and/or different conditions within the same group

Some features of HERMES (mainly the visualization tools, see section 6, DATA REPRESENTATION) require spatial information about the data. HERMES includes bi-dimensional spatial information (layouts) to simplify the representation of some of the most commonly used EEG and MEG systems, namely: i) 10-20 (21 channels) international EEG system, with 10-10 (up to 86) and 10-5 (up to 335 channels) extensions, ii) 4D Neuroimaging MAGNES 2500 WH 148 MEG system and iii) Elekta Neuroscan 306 MEG system. If possible, the system used to acquire the data is automatically detected, and the user is asked for confirmation by means of the *Layout selection* panel. This panel includes the possibility of using a variation (subset) of one of the pre-saved systems, by downsizing the number of channels as specified by the user (e.g. a subset of the 10-20 EEG system discarding the central electrodes Fz, Pz and Oz). For details about the visualization of the data/results, refer to section DATA REPRESENTATION.

## 2.1    Windowing the data

To obtain temporal resolution for the calculated indexes, a windowing parameter was included in the configuration. However, this windowing procedure should not be considered in the traditional sense of Fourier transform applications, but just as a segmentation of the given signal in smaller pieces. Thus, no window type parameter is available, as the window is always rectangular in shape.

### 2.1.1    [Windowing parameters]

The following parameters are available for all the indexes:

1) **Length of the sliding window (ms):** in case you want to calculate a time-varying version of, say, a phase synchronization (PS) index, you can divide the data in windows of a given length. This parameter fixes the length of the desired window you want to apply. RANGE: [$t_{100}$,$t_{end}$] ms; where $t_{100}$ corresponds to a minimum window of 100 samples (depending on the sampling rate, it will take a value in ms); and $t_{end}$ is the whole epoch, that is, all the data will be considered in the same window (is the same as not having windowing at all). If the data set has less than 100 samples the choice of windowing is disabled. DEFAULT: $t_{end}$ ms (no windowing)

2) **Overlapping (%):** RANGE: [0,100] %. When 100% is entered, the computation will be sliding just one sample, (note that total overlapping does not make sense). DEFAULT: 0 % (no overlapping)

3) **Windows alignment** (only available for trials): i) *With the epoch*: the windowing starts with the beginning of the data (time 0 – when the stimulus appears – does not necessarily coincide with the beginning of a new window) or ii) *With the stimulus*: the windowing starts at time 0 – when the stimulus appears – (the beginning of the data does not necessarily coincide with the beginning of a the first window). DEFAULT: *With the epoch.*

## 2.2    Calculation of the indexes

HERMES calculates different families of indexes that estimate the degree of FC and EC between signals. Each of these families contains, in turn, many indexes, each of them with a different set of configurations. The indexes to be calculated have to be selected by clicking in the checkbox next to their names. Characteristics of each index will be detailed in section 3, CONNECTIVITY MEASURES.

## 2.3    Exporting the results

Once the desired calculations have been performed in a project, the indexes may be exported as a structure stored into a MAT file or, if desired, stored in a variable named "indexes" in the workspace. This structure contains as many sub-structures as indexes were calculated, each one named after the short name of the index (i.e. COR for correlation, COH for coherence, etc.).

Calculated indexes are stored in the field named *data* (i.e. the calculated indexes using the correlation are stored in *indexes.COR.data*, where COR is the key word for correlation, as defined in section 3.1.1) in the form of a bi-dimensional cell array. This cell array contains as many columns as subjects in the project, and as many rows as conditions. Each one of these cells contains a matrix, whose dimensions are determined in the metadata of the indexes (i.e. *indexes.COR.config.dimensions* in the previous case).

Other fields in the structure contains metadata about the index configuration (*indexes.{}.config*), values for each dimension (*indexes.{}.dimensions*) or, if required, statistical significance (*indexes.{}.pval*) This structure is graphically detailed in Fig. 3.

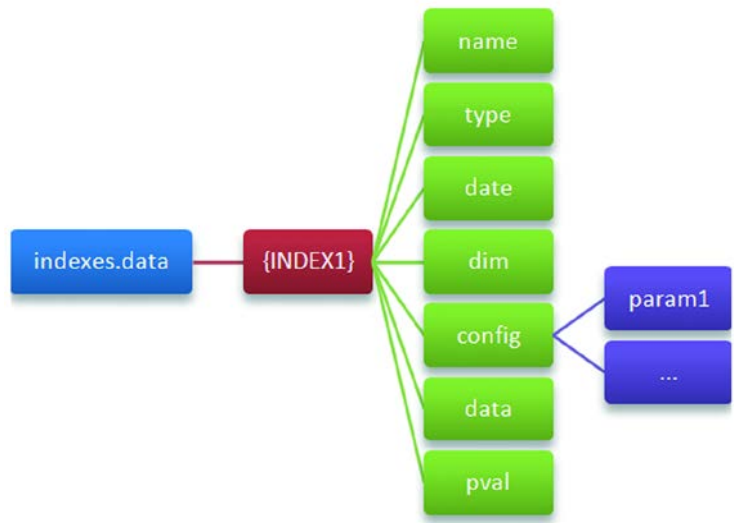

**Fig. 3 Structure of the Indexes**

A typical structure of the computed indexes {*INDEX1*}

## 2.4    Project log

Every time a set of indexes is calculated, HERMES creates a new session in the project log. This entry includes the indexes to calculate, the parameters for each family of indexes, and the time of beginning and ending of the calculation for each index. In the case of user cancelation, or if some error occurs during the execution, it is also stored in this log, for future access.

Project logs can be accessed via the "File/View project log" menu. In addition, each log session is stored as a separated file in the "<Project>/logs" directory.

## 3.  CONNECTIVITY MEASURES

As commented before, HERMES includes several types of connectivity indexes. From a conceptual point of view, they might be classified into two main groups: FC indexes, which measure statistical dependence between signals without providing any causal information, and EC indexes, which do provide such causal information. However, for the sake of clarity, we will group them in five different categories: "classical" measures (section 3.1), PS indexes (section 3.2), GS indexes (section 3.3), Granger causality-based indexes (3.4) and information theoretic indexes (section 3.5). This classification scheme is shown in Table 1.

| CM | PS | GS | GC | IT |
|----|----|----|----|----|
| COR | PLV | S | GC | MI |
| XCOR | PLI | H | DTF | TE |
| COH | WPLI | N | PDC | PMI |
| PSI | RHO | M | *AR model order (for GC)* | PTE |
| *Max lags (for XCOR)* | DPI | L | *MAR model order (for DTF and PDC)* | *Embedding Dimension* |
| *Freq Range (for PSI)* | *Center Freqs* | SL | | *Embedding Delay* |
| | *Bandwith* | *Embedding Dimension* | | *N neighbours* |
| | *Method (for DPI)* | *Embedding Delay* | | |
| | | *Theiler window ($w_1$)* | | |
| | | *N neighbors* | | |
| | | *$w_2$ and $p_{ref}$ (for SL)* | | |

**Table 1 Connectivity measures included in HERMES**

From left to right: classical measures (CM), phase synchronization measures (PS), generalized synchronization measures (GS), Granger causality-based measures (GC) and information theoretic measures (IT). *Normal font (top):* name of the indexes. *Italic font (bottom):* parameters of the indexes in each column

Henceforth we will use uppercase (X) to denote systems, italic lowercase (*x*) for signals or variables and bold lowercase (**x**) for vectors. Besides, unless stated otherwise, we will assume (without loss of generality), that signals are normalized to zero mean and unit variance.

## 3.1    Classical Measures

Classical measures include the FC linear methods most commonly used in the neuroscientific literature: Pearson's correlation coefficient (section 3.1.1), cross-correlation function (section 3.1.2) and magnitude squared coherence (section 3.1.3). These measures have the advantage of being well known and fast to compute. However, they only detect linear dependences. Besides, we have included here another measure, the Phase Slope Index (section 3.1.4) which, although recently derived (Nolte et al., 2008), is based on the classical coherency function.

### 3.1.1    *Pearson's correlation coefficient (COR)*

**DEFINITION:** Pearson's correlation coefficient measures the linear correlation in the time domain between two signals *x(t)* and *y(t)* at zero lag. For zero mean, unit variance signals it is defined as:

$$R_{xy} = \frac{1}{N} \sum_{k=1}^{N} x(k)\, y(k) \qquad (1)$$

**RANGE:** $-1 \leq R_{xy} \leq 1$. (-1): complete linear inverse correlation between the two signals, (0): no linear interdependence, (1): complete linear direct correlation between the two signals.

### 3.1.2   Cross-correlation function (XCOR)

**DEFINITION:** The cross-correlation function measures the linear correlation between two signals $x(t)$ and $y(t)$ as a function of time:

$$C_{xy}(\tau) = \frac{1}{N-\tau} \sum_{k=1}^{N-\tau} x(k+\tau)\, y(k) \qquad (2)$$

When $\tau = 0$ we recover the Pearson`s correlation coefficient (3.1.1)

**RANGE:** $-1 \leq C_{xy}(\tau) \leq 1$. (-1): complete linear inverse correlation between $x(t)$ and $y(t)$ at time delay $\tau$, (0): no linear interdependence, (1): complete linear direct correlation between $x(t)$ and $y(t)$ at time delay $\tau$.

### 3.1.3   Coherence (COH)

**DEFINITION:** The magnitude squared coherence (or simply, the coherence) measures the linear correlation between two variables $x(t)$ and $y(t)$ as a function of the frequency, $f$. It is the squared module of the coherency function ($K$), which is the ratio between the cross power spectral density, $S_{xy}(f)$, between $x(t)$ and $y(t)$, and their individual power spectral densities $S_{xx}(f)$ and $S_{yy}(f)$:

$$K_{xy}(f) = \frac{S_{xy}(f)}{\sqrt{S_{xx}(f)\, S_{yy}(f)}} \qquad (3)$$

Thus, the coherence is defined as:

$$COH_{xy}(f) = \left| K_{xy}(f) \right|^2 = \frac{\left| S_{xy}(f) \right|^2}{S_{xx}(f)\, S_{yy}(f)} \qquad (4)$$

In HERMES, we use Welch's averaged, modified periodogram method to estimate the spectrum, as we are dealing with finite data. Both the windowing of the data and the use of Welch's averaged periodogram reduce the frequency resolution of the coherence. Welch's periodogram, by default, uses segments of 2/9 times the window length, thus reducing the frequency resolution to approximately a fifth of its value.

**RANGE:** $0 \leq COH_{xy}(f) \leq 1$. (0): no linear dependence between $x(t)$ and $y(t)$ at frequency $f$. (1): correspondence between $x(t)$ and $y(t)$ at frequency $f$.

### 3.1.4 Phase Slope Index (PSI)

A key concept for the study of brain connectivity from two signals, $x(t)$ and $y(t)$, recorded from two sensor/channels is that true interactions between neural sources (as opposed to, e.g., volume conduction effects) occur with a certain time delay (Nolte et al., 2004). The existence of such a time delay in the interdependence between $x(t)$ and $y(t)$ is the conceptual basis of EC indexes such as the Granger measures covered in section 3.4. In fact, (Geweke, 1982) showed that total connectivity can be expressed as a sum of instantaneous and Granger-causal components. (Pascual-Marqui, Lehmann et al., 2011) has used this idea to derive the lagged connectivity between two signals (the one having almost pure physiological origin) as the difference between total and instantaneous connectivity. In a similar vein, (Hipp et al. 2012) recently proposed to decompose $X(t,f)$ and $Y(t,f)$ - the frequency domain versions of $x(t)$ and $y(t)$, respectively - into two components, one parallel and one orthogonal to the other signal. The component of $Y(t,f)$ parallel to $X$, $Y_{\parallel X}(t,f)$, corresponds to the part of $Y(t,f)$ that can be instantaneously and linearly predicted from $X$, and "*shares with it the co-variability in power due to measuring the same sources in two different sites*" (Hipp et al., 2012). In contrast, the component $Y(t,f)$ orthogonal to $X(t,f)$, $Y_{\perp X}(t,f) = Y(t,f) - Y_{\parallel X}(t,f)$ corresponds to that part of $Y(t,f)$ stemming from different neuronal populations to those recorded in $X(t,f)$. The power envelope correlation between $X(t,f)$ and $Y_{\perp X}(t,f)$ provides an estimation of true brain interactions as a function of time and frequency.

The aforementioned lagged connectivity component also produces a coherent phase relationship between $x(t)$ and $y(t)$ at a value different from 0 and $\pi$, which results in a complex coherency (4) with an imaginary part different from zero. Several indexes have been derived (Nolte et al., 2004, 2008; Stam, Nolte, & Daffertshofer, 2007; Vinck et al., 2011), which make use of this phase relationship to estimate the existence of true FC between $x(t)$ and $y(t)$. The current version of HERMES includes many of these indexes, such as the Phase Lag Index (PLI) and the Weighted Phase Lag Index (WPLI) (described in sections 3.2.2 and 3.2.3, respectively). The Phase Slope Index (PSI), explained henceforth, is another of such indexes. Although it can hardly be regarded as a classical measure, we cover it in this section because it is directly obtained from the complex coherency function.

**DEFINITION:** (Nolte et al., 2008) proposed a highly robust estimation of the flow direction of information between two time series by making use of the aforementioned idea. Briefly, if the speed at which different waves travel is similar, then the phase difference between the sender and the recipient of the information increases with frequency, giving rise to a positive slope of the phase spectrum. Hence, the PSI between $x(t)$ and $y(t)$ is defined as:

$$\tilde{\psi}_{xy} = \Im \left( \sum_{f \in F} K_{xy}^*(f) \, K_{xy}(f + \delta f) \right) \tag{5}$$

where $K_{xy}(f)$ is the complex coherency (as defined in (3)), $\delta f$ is the frequency resolution, $\Im(\cdot)$ denotes imaginary part and $F$ is the set of frequencies over which the slope is summed. Usually, (5) is normalized by using an estimate of its standard deviation (Di Bernardi, Nolte, & Bhattacharya, 2013; Nolte et al., 2008)

$$PSI = \psi_{xy} = \frac{\tilde{\psi}_{xy}}{std(\tilde{\psi}_{xy})} \tag{6}$$

where $std(\tilde{\psi}_{xy}) = \sqrt{k}\,\sigma$ is assessed by dividing the whole data into $k$ epochs, calculating $k$ values of (5), $\tilde{\psi}_{xy}^k$, from data with the $k_{th}$ epoch removed and then taking $\sigma$ as the standard deviation of the distribution of $\tilde{\psi}_{xy}^k$.

**RANGE:** Values of PSI > 2 indicate statistically significant time delay between $x(t)$ and $y(t)$ in the frequency range considered.

**REMARKS:** PSI *"indicates the temporal order of two signals, which is then interpreted as a driver-responder relation. For bidirectional (or unknown) coupling a finding that, e.g., A drives B does not imply that B has no impact on A. Rather, one cannot make a statement about the reverse direction."* (Nolte et al., 2008). The method has shown to outperform Granger causality in the detection of directionality in the analysis of data consisting of mixtures of independent sources.

### 3.1.5 Parameters for the classical measures

Classical measures do not need many parameters. However, for the cross-correlation we include:

1) **Max lags ($\tau$):** Is the number of lags you want to evaluate the cross-correlation. RANGE: integer between [1, Nsamples/5], to avoid edge effects (Chatfield, 1996). DEFAULT: Nsamples/20.

For the PSI, we give the option to select:

2) **Frequency band:** The beginning and the ending of the frequency band to analyse in Hz. RANGE: [0, $f_s/2$], where $f_s$ is the sampling rate. DEFAULT: All frequencies. HERMES calculates automatically from the data the number of epochs $k$ for the estimation of (6).

## 3.2   Phase synchronization indexes

PS refers to a situation when the phases of two coupled oscillators synchronizes, even though their amplitudes may remain uncorrelated (Rosenblum et al. 1996). Accordingly, for any time *t* the following equation holds:

$$\Delta\phi(t) = \left|\phi_x(t) - \phi_y(t)\right| \le cte \qquad (7)$$

which is the phase locking condition.

In experimental systems, signals are often noisy and present random phase slips of $2\pi$. Hence one has to work with the cyclic relative phase, i.e., the relative phase difference wrapped to the interval $[0,2\pi]$. It is defined as:

$$\Delta\phi_{rel}(t) = \Delta\phi(t) \bmod 2\pi \qquad (8)$$

Furthermore, in this framework, the phase locking condition (7) must be understood in a statistical sense, as the existence of a preferred value in the distribution of (8).

Before we estimate the degree of PS between two signals, some pre-processing steps are necessary, and are carried out automatically by HERMES. First, from the real-valued signals *x(t)* and *y(t)*, we obtain the corresponding analytic signals $x_{an}(t)$ and $y_{an}(t)$ (Gabor, 1946), as:

$$x_{an}(t) = x(t) + ix_H(t) = A_x(t)\, e^{i\phi_x(t)}$$
$$y_{an}(t) = y(t) + iy_H(t) = A_y(t)\, e^{i\phi_y(t)} \qquad (9)$$

where $x_H(t)$ and $y_H(t)$ are the Hilbert transforms of *x(t)* and *y(t)*, respectively[1]. Namely:

$$x_H(t) = \frac{1}{\pi} P.V. \int_{-\infty}^{\infty} \frac{x(\tau)}{t-\tau} d\tau \qquad (10)$$

where P.V. is the Cauchy's Principal Value.

Then, $A_x(t) = \sqrt{x_H(t)^2 + x(t)^2}$  and  $\phi_x(t) = arctg\dfrac{x_H(t)}{x(t)}$  are the instantaneous amplitude and phase, respectively, of $x_{an}(t)$ (analogously for $y_{an}(t)$).

---

[1] There are other ways to obtain the phase of a real-valued signal, in particular by convolving it with a complex wavelet such as the Morlet. However, these different approaches to phase extraction are known to be roughly equivalent (see, e.g., (Bruns 2004)).

In the following subsections we review the PS indexes included in HERMES, which are by far the most commonly used in neuroscientific literature: phase-locking value (PLV) (section 3.2.1), phase-lag index (section 3.2.2) , its weighted version, weighted-phase-lag index (section 3.2.3), the ρ index (section 3.2.4) and the directionality phase indexes (section 3.2.5). The parameters necessary for their calculation are detailed in section 3.2.6.

### 3.2.1    Phase Locking Value (PLV)

**DEFINITION:** The PLV makes use only of the relative phase difference (8) (Lachaux, Rodriguez, Martinerie, & Varela, 1999). It is defined as:

$$PLV = \left| \left\langle e^{i\Delta\phi_{rel}(t)} \right\rangle \right| = \left| \frac{1}{N} \sum_{n=1}^{N} e^{i\Delta\phi_{rel}(t_n)} \right| = \sqrt{\left\langle \cos\Delta\phi_{rel}(t) \right\rangle^2 + \left\langle \sin\Delta\phi_{rel}(t) \right\rangle^2}$$

(11)

where $< . >$ indicates time average.

The PLV estimates how the relative phase is distributed over the unit circle. When there is strong PS between X and Y, the relative phase occupies a small portion of the circle and the PLV is close to 1. But if the systems are not synchronized, the relative phase spreads out all over the unit circle and the PLV remains low. PLV measures the inter-trial variability of this phase difference at time *t*. PLV is also referred to in the literature as Mean Phase Coherence (Mormann, 2000) when dealing with continuous data, instead of evoked responses.

**RANGE:** $0 \leq PLV \leq 1$. (0): is very likely that the relative phase is uniformly distributed (as it would be expected, on average, for unsynchronized systems). However,  a PLV equals to zero may also occur if, e.g., this distribution has two peaks at values which differ by π. (1): if and only if the condition of strict phase locking is obeyed: phase difference is constant, and thus, complete PS is being detected.

**REMARKS:** PLV is not robust against the presence of common sources (for example, volume conduction effects (EEG and MEG) and active reference (EEG)).

### 3.2.2    Phase-Lag Index (PLI)

As commented in section 3.1.4, true interaction between two neural sources results in a coherent phase relationship between their corresponding time series at a value different from 0 and π. Instead of studying the spectrum of the coherency as in (5), this fact can be used to estimate the existence of time-lagged interdependence directly from the distribution of (8).

**DEFINITION:** This measure (Stam et al. 2007) discards phase distributions that centre around 0 mod π, in order to be robust against the presence of common sources (volume conduction and, in the case of EEG, the (possibly active) reference)

$$PLI = \left| \left\langle sign\left(\Delta\phi_{rel}(t)\right) \right\rangle \right| = \left| \frac{1}{N} \sum_{n=1}^{N} sign\left(\Delta\phi_{rel}(t_n)\right) \right|$$

(12)

**RANGE:** $0 \leq PLI \leq 1$. (0): no coupling or coupling with a phase difference cantered around 0 mod π, (1): perfect phase locking at a value of $\Delta\phi_{rel}(t)$ different from 0 mod π.

**REMARKS:** PLI is robust against the presence of common sources, but its sensitivity to noise and volume conduction is hindered by the discontinuity in this measure, as small perturbations turn phase lags into leads and vice versa, a problem that may become serious for small-magnitude synchronization effects (Vinck et al., 2011). It can be solved by using a weighted version of this index, as detailed in the next section.

### 3.2.3 Weighted Phase-Lag Index (WPLI)

As pointed out in section 3.1.4, PLI works by assessing whether the distribution of the relative phase between two signals is asymmetric around 0 or π, which ii suggests the existence of time delay between the data and therefore true interaction between the recorded sites (as opposed to e.g., volume conduction effects, which do not give rise to time delay (Nolte et al., 2004)). The problem here is that PLI, by definition, does not distinguish whether a value of the relative phase is close to zero or not, the only things that matters is whether it is positive (producing a +1) or negative (-1). Thus, in the case of noisy signals, where values of the relative phase are close to zero may change from lead (+1) to lag (-1) only due to the presence of noise, PLI is biased and loses some ability to detect changes in PS specially in the case of weak coupling (Vinck et al., 2011). This problem can be solved if this discontinuity in the effect of the relative phase (which varies from +1 to -1) on PLI is eliminated by taking into account not only the phase, but also the amplitude of the imaginary component of the cross spectrum. In this way, relative phases corresponding to a small imaginary cross-spectrum have a lower effect on the corresponding PS index, which is defined henceforth.

**DEFINITION:** A weighted version of PLI, termed WPLI, has been recently developed to tackle the problems of PLI indicated above (Vinck et al., 2011). Its relation with PLI lies in the fact that WPLI weights $sign(\mathfrak{I}(X))$ by $/|\mathfrak{I}(X)/$, where $\mathfrak{I}(X)$ is the imaginary component of the cross-spectrum between $x(t)$ and $y(t)$:

$$WPLI = \frac{\left| \left\langle \mathfrak{I}(X) \right\rangle \right|}{\left\langle \left| \mathfrak{I}(X) \right| \right\rangle} = \frac{\left| \left\langle \left| \mathfrak{I}(X) \right| sign\left(\mathfrak{I}(X)\right) \right\rangle \right|}{\left\langle \left| \mathfrak{I}(X) \right| \right\rangle}$$

(13)

**RANGE:** $0 \leq \text{WPLI} \leq 1$. (0): no synchronization, (1): synchronization: P{ *sign( $\mathfrak{J}(X)$ )=1* }=1 or P{ *sign( $\mathfrak{J}(X)$ )=-1* }=1, where P{.} denotes probability.

**REMARKS**: Differently from PLI, in WPLI the contribution of the observed phase leads and lags is weighted by the magnitude of the imaginary component of the cross-spectrum, which results in the latter index presenting "*reduced sensitivity to uncorrelated noise sources and increased statistical power to detect changes in PS*" (Vinck et al., 2011). Note that the WPLI, contrary to the rest of the PS indexes, mixes both phase and amplitude information. But we have included it here because it is directly related to the PLI, and thus we believe it is better categorized as a PS index. We recommend the interested readers to peruse (Vinck et al., 2011) for a thorough comparison of the properties of the coherence, PLV, PLI and WPLI (see, for instance the very informative table 1 within this reference).

### 3.2.4   $\rho$ index (RHO)

**DEFINITION:** This index is based on Shannon entropy (Tass et al., 1998). It quantifies the deviation of the distribution of the cyclic relative phase from the uniform distribution, approximating the probability density by the relative frequencies from the histogram of relative phases. It is defined as:

$$\rho = \frac{S_{max} - S}{S_{max}}$$

(14)

where $S_{max}$ is the maximal entropy (that of uniform distribution), i.e., the logarithm of the number of bins in the histogram, and $S$ is the entropy of the distribution of $\Delta\phi_{rel}(t)$:

$$S = -\sum_{k=1}^{N} p_k \ln(p_k)$$

(15)

where, $p_k$ is the probability of finding $\Delta\phi_{rel}(t)$ in the *k-th* bin.

**RANGE:** $0 \leq \rho \leq 1$. (0): uniform distribution (no synchronization), (1): Dirac-like distribution (perfect synchronization).

### 3.2.5   *Directionality Phase Indexes (DPI)*

All the PS indexes described hitherto work by analysing the distribution of the relative phase in different ways. Yet it is also possible to derive directionality PS indexes by analysing the temporal evolution of the phase derivative (Rosenblum et al, 2002; Rosenblum & Pikovsky, 2001). The underlying idea is that if two self-sustained oscillators *x(t)* and *y(t)* are weakly coupled, the increment of their phases depends only on the phases themselves, without any effect of the amplitudes. Thus, such increment can be modelled by means of periodic functions of both phases, and the existence of directionality in the PS between the oscillators can be assessed by the parameters of these functions, as explained henceforth.

**DEFINITION:** Two different model-based DPI are covered in HERMES:

**Evolution map approach (EMA)** (Rosenblum & Pikovsky, 2001)

Let us consider increments of phases during a fixed time interval τ:

$$\Delta_x(k) = \phi_x(t_k + \tau) - \phi_x(t_k) = \omega_x \tau + \mathcal{F}_x\Big(\phi_x(t_k), \phi_y(t_k)\Big) + \eta_x(t_k) \qquad (16)$$

where the phases are unwrapped (i.e., not reduced to the interval [0,2π)). Here, $\omega_x$ is the frequency of *x(t)*, $\eta_x$ $(t_k)$ represents the noise component of the phase increment (i.e., the error in the model) and $t_k = \delta t k$, where δt is the sampling interval. The function $F_x$, which represents the deterministic part of the model, can be estimated from the time series $\Delta_x(k)$ and $\phi_x(t_k)$ by using as a natural probe function a finite Fourier series (Rosenblum et al., 2002):

$$\mathcal{F}_x = \sum_{m,l} A_{m,l} e^{i(m\phi_x + l\phi_2)} \qquad (17)$$

From this function, one computes the cross dependence of the phase dynamics of *x(t)* on the phase of *y(t)* as:

$$c_x{}^2 = \iint_0^{2\pi} \left(\frac{\partial F_x}{\partial \phi_y}\right)^2 d\phi_x \, d\phi_y \qquad (18)$$

Then, by defining $\Delta_y$ $(k)$ in complete analogy with (16) and proceeding in the same way, we can also obtain $c_y{}^2$, which estimates the dependence of $\Delta_y$ $(k)$ on $\phi_x$ $(t_k)$.

Finally, a directionality index of PS can be computed from $c_x{}^2$ and $c_y{}^2$:

$$d^{xy} = \frac{c_x - c_y}{c_x + c_y} \qquad (19)$$

**RANGE:** $-1 \leq d^{xy} \leq 1$. (1): unidirectional coupling (x->y), (-1): opposite case (y->x), $(-1 < d^{xy} < 1)$ intermediate values correspond to bidirectional coupling.

**Instantaneous Period Approach (IPA)** (Rosenblum et al., 2002)**:**

Instead of studying phase increments of the weakly coupled, self-sustained oscillators, we can also look at the evolution of their *instantaneous periods:*

$$T_x(k) = T_x^0(k) + \Theta_x(\phi_x(t_k), \phi_y(t_k)) + \eta_x(t_k) \qquad (20)$$

where $T_x^0(k)$ is the mean period of *x(t)*, and $\eta_x(t_k)$ and $t_k$ are defined as in EMA above. Again, the deterministic part of the dependence $\Theta_x$ can be estimated by fitting a Fourier time series and the dependence of (20) on $\phi_y(t_k)$, $c_x^2$, can be calculated from $\Theta_x$ in complete analogy to (18). We then proceed likewise to obtain $c_y^2$ from $\Theta_y$ by modelling $T_y(k)$ as in (20). Finally, a second directionality index is defined:

$$r^{xy} = \frac{c_x - c_y}{c_x + c_y} \qquad (21)$$

**RANGE:** $-1 \leq r^{xy} \leq 1$. (1): unidirectional coupling (x->y), (-1): opposite case (y->x), ($-1 < r^{xy} < 1$) intermediate values correspond to bidirectional coupling.

**REMARKS:** Several remarks are in order. First and foremost, both EMA and IPA are based on the assumption that the coupling between the oscillators is weak. If it is not case (e.g., if the value of, say, PLV between *x(t)* and *y(t)* is high), then the phase increments (and the instantaneous period) are also influenced by amplitudes, (16) and (20) no longer hold and either of the indexes are meaningless.

Second, although both model-based PS indexes (19) and (21) derived respectively from EMA and IPA may seem at first sight equivalent, the latter one reflects not only asymmetry in coupling coefficients and functions (as the former one does) but also asymmetry in natural frequencies, so it must be used with care if the frequencies of both signals are different (see Rosenblum et al. (2002) for a detailed technical discussion of this difference). Moreover, once the values of *m* and *l* in the Fourier series modelling *F* and $\Theta$ are fixed (we take *m=l=3* following Rosenblum et al., 2002), EMA index depends on the choice of $\tau$, whereas $r^{xy}$ is parameter-free.

Finally, it has been shown that for short noisy time series the estimation of both indexes is biased (Smirnov & Bezruchko, 2003), a problem that has to be taken into account when dealing with this kind of data.

### 3.2.6 *Parameters for the PS indexes*

For PS indexes the following parameters can be modified:

1) **Central band frequencies (Hz):** You can type all the specific central band frequencies you want to analyse. Or, if you prefer, you can introduce the initial and the final frequency as well as the steps between frequencies ':' (e.g.: '5:10:40', would be equivalent to '5 15 25 35'). RANGE: [0, $f_s/2$] Hz, where $f_s$ is the sampling frequency of the data. The frequency $f_s/2$ is the Nyquist frequency, the maximum you can analyse (see, e.g., (Bendat & Piersol, 2010)). To analyse a frequency band which includes the 0, a low pass filter is computed; when the band includes $f_s/2$, a high pass filter is applied. DEFAULT: $f_s/4$ Hz (centre of the Nyquist band).

2) **Bandwidth (Hz):** the spectral windows are in the interval ($f$-$b_w/2$, $f$+$b_w/2$), where '$b_w$' is the window's bandwidth. RANGE: [4, $f_s/2$] Hz. DEFAULT: 4 Hz.

For the model-based DPIs, the method can also be selected:

3) **Method**: one has to choose whether '*EMA*' or '*IPA*', are used so that the corresponding indexes $d^{xy}$ or $r^{xy}$ are calculated, respectively. DEFAULT: '*EMA*'.

When selecting EMA another special parameter is automatically computed:

4) **τ (in samples)**: Following Rosenblum et al. (2002), it is set to τ = min ($T_1,T_2$), where $T_1$ and $T_2$ are the periods of oscillations of *x(t)* and *y(t)*, respectively.

## 3.3 Generalized synchronization indexes

The concept of GS refers to a situation where the states of a dynamical (sub)system Y are a function of those of another one X, namely Y=F(X). Even though we deal with experimental (sub)systems, where the function F (and possible also the state equation of each system) are unknown, and where F may be complex and changing over time, the existence of GS between X and Y has an important consequence: similar states tend to occur at similar times in both subsystems. In terms of the signals that the (sub)systems generate, this means that if the temporal patterns in *x(t)* at times $t_i$ and $t_j$ are similar, likewise the patterns in *y(t)* at these same times will be similar, something that can be quantified with different indexes, as detailed henceforth.

### 3.3.1 First processing steps

From two different simultaneously recorded time series $x(t)=(x_1,x_2,...,x_N)$ and $y(t)=(y_1,y_2,...,y_N)$, delay phase-space vectors can be constructed with embedding dimension $d$ and time delay τ (Takens, 1981)[2]:

$$\mathbf{x_n} = \big( x(n), x(n-\tau), \ldots, x(n-(d-1)\tau) \big)$$
$$\mathbf{y_n} = \big( y(n), y(n-\tau), \ldots, y(n-(d-1)\tau) \big) \tag{22}$$

Let $\mathbf{r_{n,j}}$ and $\mathbf{s_{n,j}}$ , j=1,…,k, denote the time indices of the $k$ nearest neighbours of $\mathbf{x_n}$ and $\mathbf{y_n}$ , respectively.

---

[2] Although this is normally the case, $d$ and $\tau$ should not necessarily be the same for *x(t)* and *y(t),* but in the following, we assume that they are indeed equal for both signals. However, we will see later how it is possible to take a different value of $d$ for *x* and *y* (see section 3.5.3).

The mean Euclidean distance of $\mathbf{x_n}$ to its $k$ nearest neighbours is:

$$R_n^{(k)}(X) = \frac{1}{k} \sum_{j=1}^{k} \left| \mathbf{x_n} - \mathbf{x_{r_{n,j}}} \right|^2 \qquad (23)$$

Additionally, the Y-conditioned mean squared Euclidean distance can be obtained, by replacing the nearest neighbours by the equal time partners of the nearest neighbours of $\mathbf{y_n}$

$$R_n^{(k)}(X \mid Y) = \frac{1}{k} \sum_{j=1}^{k} \left| \mathbf{x_n} - \mathbf{x_{s_{n,j}}} \right|^2 \qquad (24)$$

Finally, the radius of the reconstructed phase space of X is defined as:

$$R_n(X) = \frac{1}{N-1} \sum_{\substack{j=1 \\ j \neq n}}^{k} \left| \mathbf{x_n} - \mathbf{x_j} \right|^2 \qquad (25)$$

As we will see in the following sections, GS-based indexes make use of these distances in their definitions (Rulkov, Sushchik, & Tsimring, 1995). In HERMES, the following GS indexes can be calculated: S, H, N, M, L and synchronization likelihood (SL).

### 3.3.2   S Index

**DEFINITION:** The S index (Arnhold, 1999) is defined as:

$$S^{(k)}(X \mid Y) = \frac{1}{N} \sum_{n=1}^{N} \frac{R_n^{(k)}(X)}{R_n^{(k)}(X \mid Y)} \qquad (26)$$

**RANGE:** $0 < S\ (X|Y) \leq 1$. $(0^+)$: independence between X and Y, (1): complete generalized synchronization

**REMARKS:** Not very robust against noise and signal length.

### 3.3.3   H Index

**DEFINITION:** H index (Arnhold, 1999) is defined as:

$$H^{(k)}(X \mid Y) = \frac{1}{N} \sum_{n=1}^{N} \frac{\log(R_n(X))}{R_n^{(k)}(X \mid Y)}$$

(27)

**RANGE:** $0 < H(X|Y) < \infty$. (0): suggests (but does not prove it) that X and Y are independent. If X and Y are completely independent, it is 0. (>0): nearness in Y implies also nearness in X for equal time partners,

**REMARKS:** H is more robust against noise and easier to interpret than S, but with the drawback that it is not normalized.

### 3.3.4    N Index

**DEFINITION:** the *N* index (Quiroga et al. 2002) is defined as:

$$N^{(k)}(X \mid Y) = \frac{1}{N} \sum_{n=1}^{N} \frac{R_n(X) - R_n^{(k)}(X \mid Y)}{R_n(X)}$$

(28)

**RANGE:** $0 \leq N(X|Y) < 1$. (0): X and Y are independent, (1): X and Y are synchronized.

**REMARKS:** *N*(X|Y) is normalized (but as in the case of H, it can be slightly negative) and in principle more robust than S. It reaches its maximum value of 1, only if $R_n^{(k)}(X|Y) = 0$, which does not happen even if X and Y are identically synchronized (except for periodic signals). This small drawback was corrected in the following index.

### 3.3.5    M Index

**DEFINITION:** (Andrzejak et al. 2003):Another way of normalizing ratios:

$$M^{(k)}(X \mid Y) = \frac{1}{N} \sum_{n=1}^{N} \frac{R_n(X) - R_n^{(k)}(X \mid Y)}{R_n(X) - R_n^{(k)}(X)}$$

(29)

**RANGE:** $0 \leq M(X|Y) \leq 1$. (0): X and Y are independent, (1): X and Y are fully synchronized.

### 3.3.6    L Index

**DEFINITION:** This is a GS index where distances are calculated between ranks of the vectors (i.e., they are *normalized*) (Chicharro & Andrzejak 2009):

$$L^{(k)}(X \mid Y) = \frac{1}{N} \sum_{n=1}^{N} \frac{G_n(X) - G_n^{(k)}(X \mid Y)}{G_n(X) - G_n^{(k)}(X)} \tag{30}$$

Average rank: $G_n(X) = \dfrac{N}{2}$, Minimal average rank: $G_n^{(k)}(X) = \dfrac{k+1}{2}$

The Y-conditioned average rank is: $G_n^{(k)}(X \mid Y) = \dfrac{1}{k} \sum_{j=1}^{k} g_{n,s_{nj}}$, where $g_{i,j}$ are the ranks that the distance between $\mathbf{x_i}$ and $\mathbf{x_j}$ occupy in a sorted ascending list of distances between $\mathbf{x_i}$ and all $\mathbf{x_{j \neq i}}$.

**RANGE:** $0 \leq L\,(X|Y) \leq 1$. (0): X and Y are independent, (1): X and Y are synchronized.

**REMARKS:** L performs better for the detection of directionality in the interdependence than the rest of the GS-bases indexes.

Although unlikely, it is not impossible (mainly in those cases where the parameter *k* is too low, see section 3.3.8 and/or the coupling is rather weak) that nearest neighbours of the reference vectors in X correspond to vectors in Y whose average distance to the simultaneous reference vectors in this latter state space is greater than expected for randomly picked vectors. In this case, all the GS indexes described hitherto will be slightly negative, but only because the *k* mutual neighbours are a biased (i.e., too small) sample of the reconstructed attractor. For all practical purposes, however, this is equivalent to the indexes being equal to zero (no GS), and HERMES actually set them to zero if the user decides to, at the same time producing a warning for him/her to be aware that the value of *k* should be increased.

### 3.3.7   *Synchronization Likelihood (SL)*

**DEFINITION:** Synchronization likelihood (SL) (Stam & van Dijk, 2002) is arguably the most popular index to estimate GS in neurophysiological data. This index, which is closely related to the concept of generalized mutual information (Buzug et al., 1994), relies on the detection of simultaneously occurring patterns, which can be complex and widely different across signals. Contrary to all the GS indexes described hitherto, which assess the existence of connectivity between only two signals *x(t)* and *y(t)*), SL is truly multivariate, as it gives a normalized estimate of the dynamical interdependencies between *M* ($\geq$2) time series $x_1(t),..., x_M(t)$. Thus, the corresponding *d*-dimensional delayed vectors at time *n* are defined as:

$$\mathbf{x_{1,n}} = \left( x_1(n), x_1(n-\tau), \ldots, x_1(n-(d-1)\tau) \right)$$
$$\mathbf{x_{2,n}} = \left( x_2(n), x_2(n-\tau), \ldots, x_2(n-(d-1)\tau) \right)$$
$$\vdots \tag{31}$$
$$\mathbf{x_{M,n}} = \left( x_M(n), x_M(n-\tau), \ldots, x_M(n-(d-1)\tau) \right)$$

where $\tau$ is the delay time. Of course, the expression above reduces to (22) for the bivariate case, with $x_1(t) = x(t)$ and $x_2(t) = y(t)$.

The probability that two embedded vectors from signal $x_m(t)$ $(m=1,..,M)$ are closer to each other than a given distance $\varepsilon$ at time $n$ is given by:

$$P_{m,n}^{\varepsilon} = \frac{1}{2(w_2 - w_1)} \sum_{\substack{j=1 \\ w_1 < |n-j| < w_2}}^{N} \Theta(\varepsilon - |\mathbf{x}_{m,n} - \mathbf{x}_{m,j}|) \qquad (32)$$

where $\Theta$ is the Heaviside step function ($\Theta(\mathbf{x}) = 1$ if $\mathbf{x} > 0$, 0 otherwise), $w_1$ is the Theiler window, used to avoid autocorrelation effects on the calculations, and should be at least of the order of the autocorrelation time (Theiler, 1986); and $w_2$ is a window that sharpens the time resolution of the synchronization measure and is chosen such that $w_1 << w_2 << N$ (see (Montez et al., 2006) for details on how to calculate $w_2$).

Now for each of the $M$ signals considered and each time $n$, the critical distance $\varepsilon_{m,n}$ is determined for which $P_{m,n}^{\varepsilon_{m,n}} = p_{ref} << 1$, where $p_{ref}$ denotes the percentage of reconstructed state vectors in $x_m(t)$ close enough to $\mathbf{x}_{m,n}$ to be regarded as being dynamically equivalent to them. We now determine for each discrete time pair $(n,j)$ within the time window $w_1 < |n - j| < w_2$ the number of channels $H_{n,j}$ where the embedded vectors $\mathbf{x}_{m,n}$ and $\mathbf{x}_{m,j}$ will be closer together than this critical distance $\varepsilon_{m,n}$:

$$H_{n,j} = \sum_{m=1}^{M} \Theta(\varepsilon_{m,n} - |\mathbf{x}_{m,n} - \mathbf{x}_{m,j}|) \qquad (33)$$

This number lies in the range between 0 and M, and reflects how many of the embedded signals 'resemble' each other. Then, we define a *synchronization likelihood* $S_{m,n,j}$ for each signal *m* and discrete time pair $(n,j)$ as:

$$if \; |\mathbf{x}_{m,n} - \mathbf{x}_{m,j}| < \varepsilon_n : \quad S_{m,n,j} = \frac{H_{n,j} - 1}{M - 1}$$
$$if \; |\mathbf{x}_{m,n} - \mathbf{x}_{m,j}| \geq \varepsilon_n : \quad S_{m,n,j} = 0 \qquad (34)$$

By averaging over all $j$, we finally obtain the synchronization likelihood, $SL_{m,n}$:

$$SL_{m,n} = \frac{1}{2(w_2 - w_1)} \sum_{\substack{j=1 \\ w_1 < |n-j| < w_2}}^{N} S_{m,n,j} \qquad (35)$$

$SL_{m,n}$ describes how strongly channel $x_m(t)$ at time $n$ is synchronized to all the other M − 1 channels.

We can finally get a value of the SL for the whole time interval considered ($SL_m$) by averaging $SL_{m,n}$ for all $n$.

**RANGE:** $p_{ref} \leq \text{SL} \leq 1$. ($p_{ref}$): all M time series are uncorrelated, (1): maximal synchronization of all M time series. The value of $p_{ref}$ can be set at an arbitrarily low level, and does not depend on the properties of the time series, nor is it influenced by the embedding parameters.

**REMARKS:** More detailed information about SL and its use in filtered signals can be found in (Montez et al., 2006; Stam & van Dijk, 2002).

### 3.3.8    *Parameters for the GS indexes*

To obtain reliable results for the GS measures, we have to choose the correct parameters and, in particular, proper values for the embedding dimension and the delay time. For those interested in getting further insight into this issue, we recommend the book from (Kantz & Schreiber 2004).

1) **Embedding dimension ($d$):** There are different ways to estimate a proper value of the embedding dimension for reconstruction, such as the false nearest neighbours method (Kennel, Brown, & Abarbanel, 1992). RANGE: integer between [2,10]. DEFAULT: The value $d_c$ for which the percentage of false nearest neighbours falls below 10%.

2) **Embedding delay ($\tau$):** RANGE: integer between [1, *0.8Nsamples/(d-1)*], where 1 is for consecutive samples and the upper bound of the range is set to *0.8Nsamples/(d-1)* to guarantee that the number of vectors is at least 20% of the number of samples. DEFAULT: autocorrelation time (act) of the signal, i.e., the time when the envelope of the autocorrelation function decreases to 1/e (0.32)

3) **Theiler window ($w_1$):** to exclude autocorrelation effects from the density estimation, discarding for the nearest-neighbour search those samples which are closer in time to a reference point than a given lapse. RANGE: [$\tau$, $2\tau$]. DEFAULT: $\tau$ (delay time) (Theiler, 1986).

4) **Number of nearest neighbours ($k$):** RANGE: [$d$, $2d$]. This is set to prevent a bias in the value of the indexes (Pereda et al., 2001), as for instance the one described at the end of section 3.3.6. DEFAULT: $d+1$.

For SL, parameters $d$, $\tau$ and $w_1$ have the same ranges and default values as for the rest of GS indexes[3]. Thus, the only parameters specific from SL are $w_2$ and $p_{ref}$.

1) **$p_{ref}$ :** RANGE: [0.01, 0.5]. DEFAULT: 0.05. **$p_{ref}$** = 0.05 means that five per cent of the vectors $\mathbf{x_{m,j}}$ will be considered recurrences of $\mathbf{x_{m,n}}$

---

[3] In case the signals are narrowband (i.e., they have been band-pass filtered to calculate the SL in a given frequency band), it is advisable to take into account the recommendations included in (Montez et al., 2006) for the values of all the parameters of this index.

2) **$w_2$:** $w_2 = n_{rec}/p_{ref} + w_1 - 1$, where $n_{rec}$ is the number of recurrences, which is set to $n_{rec}=10$. A low value of $p_{ref}$ should only be used for long enough time series, as it may give rise to a high value of $w_2$.

As a final comment on the GS indexes, note that their estimation entails the calculation and sorting of distances in *d*-dimensional spaces, a procedure that, depending on data length and number of channels/sensors, maybe rather time consuming and demanding in terms of memory usage even for a modern computer. Although we have done our best to optimize the code for these indexes[4], it is advisable to check the progress bar drawn during the calculations, which gives a fair estimation of the time needed for the computations. It may help to determine whether it is advisable to change the calculation parameters and/or reduce data length to get reasonable computational times.

## 3.4 Granger causality measures

### 3.4.1 Classical Linear Granger Causality (GC)

**DEFINITION:** For two simultaneously measured signals *x(t)* and *y(t)*, if one can predict the first signal better by incorporating the past information from the second signal than using only information from the first one, then the second signal can be called causal to the first one (Wiener, 1956). It was the Nobel Prize laureate Clive Granger who gave a mathematical formulation of this concept (Granger, 1969) by arguing that when *x* is influencing *y*, then if you add past values of *x(t)* to the regression of *y(t)*, an improvement on the prediction will be obtained.

For the univariate autoregressive model (AR), we have:

$$x(n) = \sum_{k=1}^{P} a_{x,k} x(n-k) + u_x(n)$$
$$y(n) = \sum_{k=1}^{P} a_{y,k} y(n-k) + u_y(n)$$

(36)

where $a_{i,j}$ are the model parameters (coefficients usually estimated by least square method), *p* is the order of the AR model and $u_i$ are the residuals associated to the model. Here, the prediction of each signal (*x* and *y*) is performed only by its own past ($\overline{x}$ and $\overline{y}$ respectively). The variances of the residuals are denoted by:

---

[4] To get an idea of typical computational times, we have tested the current version of HERMES in Matlab® 2012b 64-bit (Windows® 8 Pro operating system) running on an Intel® Core™ I7-3770K CPU @ 3.5 GHz with 16 Gb RAM, and it takes less than one minute per trial to calculate all the GS indexes from section 3.3.2 to 3.3.6. (512 samples, 128 MEG sensors, *d*=6, $\tau$=7, *k*=10, $w_1$=8).

$$V_{x|\bar{x}} = \text{var}(u_x)$$
$$V_{y|\bar{y}} = \text{var}(u_y)$$

(37)

And for the bivariate AR:

$$x(n) = \sum_{k=1}^{P} a_{x|x,k} x(n-k) + \sum_{k=1}^{P} a_{x|y,k} y(n-k) + u_{xy}(n)$$
$$y(n) = \sum_{k=1}^{P} a_{y|x,k} x(n-k) + \sum_{k=1}^{P} a_{y|y,k} y(n-k) + u_{yx}(n)$$

(38)

where the residuals now depend on the past values of both signals and their variances are:

$$V_{x|\bar{x},\bar{y}} = \text{var}(u_{xy})$$
$$V_{y|\bar{x},\bar{y}} = \text{var}(u_{yx})$$

(39)

where *var(.)* is the variance over time and *x|x,y* is the prediction of *x(t)* by the past samples of values of *x(t)* and *y(t)*.

Therefore, Granger causality (GC) from *y* to *x* (predicting *x* from *y*) is:

$$GC_{y \to x} = \ln\left(\frac{V_{x|\bar{x}}}{V_{x|\bar{x},\bar{y}}}\right)$$

(40)

**RANGE:** $0 \leq GC_{Y \to X} < \infty$. (0): the past of *y(t)* does not improve the prediction of *x(t)*: $V_{x|\bar{x}} \approx V_{x|\bar{x},\bar{y}}$. (>0): the past of Y improves the prediction of X: $V_{x|\bar{x}} \gg V_{x|\bar{x},\bar{y}}$ (*y* G-causes *x*)

**REMARKS:** GC has the advantage of been asymmetric; therefore, it is able to detect effective connectivity. However, it is a linear parametric method, so it depends on the autoregressive model of order *p*. For those readers interested in further exploring GC and its different variants, we recommend the excellent GCCA toolbox (Seth, 2010).

### 3.4.2   *Partial Directed Coherence (PDC)*

**DEFINITION:** The PDC provides a frequency domain measure based in Granger causality (Baccalá & Sameshima, 2001; Sameshima & Baccalá, 1999). It is based on modelling time series by multivariate autoregressive (MAR) processes. Consider a MAR process of order *p* with dimension *M* (i.e., *M* signals simultaneously measured, $x_1(t),..., x_M(t)$):

$$\begin{pmatrix} x_1(k) \\ \vdots \\ x_M(k) \end{pmatrix} = \sum_{r=1}^{p} A_r \begin{pmatrix} x_1(k-r) \\ \vdots \\ x_M(k-r) \end{pmatrix} + \begin{pmatrix} \varepsilon_1(k) \\ \vdots \\ \varepsilon_M(k) \end{pmatrix} \qquad (41)$$

where $A_1, A_2, ..., A_p$ are *MxM* coefficient matrices, and $\varepsilon_i(k)$ are independent Gaussian white noises with covariance matrix $\Sigma$.

We can get the frequency domain version of (41) by computing the power spectral density matrix:

$$S(f) = \mathbf{H}(f) \sum \mathbf{H}^H(f) \qquad (42)$$

where (.)$^H$ is the Hermitian transpose, H is the transfer function matrix: $H(f) = \bar{A}^{-1}(f) = [I-A(f)]^{-1}$, $A(f)$ is the Fourier transform of the coefficients and $\bar{A}(f) = [\bar{a}_1(f)\ \bar{a}_2(f)...\ \bar{a}_M(f)]$, with $\bar{a}_{ij}(f)$ being the *i,j* th element of $\bar{A}(f)$.

Then, the PDC from signal *j* to signal *i* is given by:

$$PDC(f) = \pi_{ij}(f) = \frac{\overline{a}_{ij}(f)}{\sqrt{\mathbf{a}_j^H(f)\overline{\mathbf{a}}_j(f)}} \qquad (43)$$

$\pi_{ij}$(f) represents the relative coupling strength of the interaction of a given source (signal *j*), with regard to some signal *i*, as compared to all of the *j*'s connections to other signals. We have that $\Sigma_i|\pi_{ij}(f)|^2=1$, for all $1 \leqq j \leqq M$. For $i = j$, the PDC, $\pi_{ii}(f)$, represents how much of $X_i$'s own past is not explained by other signals.

**RANGE:** $0 \leq |\pi_{ij}(f)|^2 \leq 1$. (0): no coupling, (1): complete coupling.

### 3.4.3   Direct Transfer Function (DTF)

**DEFINITION:** The DTF is defined similarly to the PDC (Kaminski & Blinowska, 1991)

$$DTF(f) = \vartheta_{ij}(f) = \frac{H_{ij}(f)}{\sqrt{\mathbf{h}_j^H(f)\mathbf{h}_j(f)}} \qquad (44)$$

DTF uses the elements of the transfer function matrix H, whereas PDC uses those of Ā.

**RANGE:** $0 \leq DTF \leq 1$. (0): no coupling, (1): complete coupling

**REMARKS:** DTF calculation does involve matrix inversion, so PDC is computationally more efficient and more robust than DTF. Furthermore, PDC is normalized with respect to the total inflow of information, but DTF is normalized with respect to the total outflow of the information. Due to matrix inversion, PDC is able to ignore indirect influences, and detect only direct ones.

### 3.4.4   *Parameters for the GC measures*

1) **Order of the AR model** (for GC)**:** RANGE: [3, *Nsamples-1*]. DEFAULT: $p = \min(p_1, p_2)$, where $p_1$ and $p_2$ are the values obtained by applying the Akaike (Akaike, 1974) and the Bayesian Information Criterion (Schwarz, 1978), respectively.

2) **Order of the MAR model** (for PDC and DTF): RANGE: [3, *(Nsamples*Ntrials-1)/ (Nchannels+Ntrials) - 1]*. DEFAULT: $p = p_{AR} / k$, where $p_{AR}$ is the AR model default order and $k$ is a number between 1 and 5 proportional to the mean correlation of the data (5 for highly correlated data and 1 for weakly correlated data).

The coefficients of the MAR model of the desired order are estimated by means of a simplified version of the ARfit toolbox (Neumaier & Schneider, 2001; Schneider & Neumaier 2001). The number of points for the FFT (in the case of the PDC and the DTF) is equal to the next power of 2 greater than the window length.

**REMARKS:** The successful estimation of PDC or DTF depends primarily on the reliability of the fitted MAR model (optimal model order and epoch length). If the model order is too low, the model will not capture the essential dynamics of the data set, whereas if the model order is too high, it will also capture the unwanted component (i.e., noise), leading to over-fitting and instability. MAR model assumes that the underlying process is stationary, but neurophysiological and cognitive events are themselves transient and may rapidly change their states, insomuch as the neural signals are often non-stationary (Pereda et al. 2005). Theoretically, the span of the chosen window can be as short as $p + 1$ data points, where $p$ is the model order. Practically, such a short window would be impossible to achieve for a single realization of the multivariate data set. As a result, a balance has to be maintained between time resolution (limited by stationarity) and the statistical properties of the fitted model. As a rule of thumb, the window length should possess a few times more data points than the number of estimated model parameters. An alternative solution has been offered by (Ding et al., 2000), where the collection of neural signals from successive trials is treated as an ensemble of realizations of a non-stationary stochastic process with locally stationary segments.

## 3.5   **Information theoretic measures**

Information theory is mainly based on a measure that quantifies the information of a discrete random variable X: its Shannon entropy (Shannon & Weaver, 1949; Shannon, 1948) given by:

$$H(X) = -\sum_x p(x) \log_2 p(x) \qquad (45)$$

Note that (45) differs from (15) only in the basis of the logarithm, but in both cases entropy quantifies the reduction in the uncertainty about a variable when it is measured (or equivalently, the average information content missed when one does not know this value).

In the following we work with continuous random variable X, so we compute the differential entropy, which is defined by:

$$H(X) = -\int_{\mathbb{R}^d} f(x) \log\big(f(x)\big)\, dx \qquad (46)$$

where $f: \mathbb{R}^d \rightarrow \mathbb{R}$ is the probability density function of X.

For the estimation of the Shannon differential entropy, the Kozachenko-Leonenko (KL) estimator is used. KL estimator is a non-parametric estimator based on $k^{th}$ nearest neighbours of a sample set (Kozachenko & Leonenko, 1987).

In HERMES we include two indexes that estimate interdependence between two signals based on this concept: mutual information (MI), (section 3.5.1), and transfer entropy (section 3.5.3) (Schreiber, 2000), which derives from a Wiener causal measure (Wiener, 1956) within the framework of information theory. Additionally, the toolbox includes their corresponding partialized versions, namely, partial mutual information (section 3.5.2) and partial transfer entropy (section 3.5.4).

### 3.5.1   *Mutual Information(MI)*

**DEFINITION:** Mutual information quantifies the amount of information that can be obtained about a random variable by observing another.

$$MI_{xy} = \sum_i p(x, y) \log \frac{p(x, y)}{p(x)\, p(y)} \qquad (47)$$

It measures the amount of information shared by $x$ and $y$. Its importance lies in the fact that if $MI_{xy}=0 \leftrightarrow x$ and $y$ are independent. HERMES estimates (47) as a combination of entropies, $MI_{XY} = H(X) + H(Y) - H(X,Y)$, where $H$ is the differential entropy (45).

**RANGE:** $0 \leq MI_{xy} < \infty$. (0): $x$ and $y$ are independent, (>0): $x$ and $y$ are dependent

**REMARKS:** The main strength of $MI_{xy}$ is that it detects (if any) high order correlations, as it is based on probability distributions. Therefore, it does not rely on any specific model of the data. However, it does not identify causal relationships, due to its lack of directional information[5].

Recently an optimized version of mutual information termed maximal information coefficient (MIC) has been derived, (Reshef et al., 2011): for each pair of samples $(x_n, y_n)$ from signals $x(t)$ and $y(t)$, the MIC algorithm works by finding the *n-by-m* grid with the highest induced $MI_{xy}$. It then compiles a matrix that stores, for each resolution, the best grid at that resolution and its normalized score. MIC corresponds to the maximum of these normalized scores for the range of grids considered. This index, which has been regarded by some researchers as "*a correlation for the 21st century*" (Speed, 2011), has been included in a toolbox for different programming languages (including Matlab®) published early this year (Albanese et al., 2013).

### 3.5.2   *Partial Mutual Information (PMI)*

As commented in the last section, mutual information estimates the amount of information shared between *x* and *y*. However, it does not give any clue as to whether this shared information is the result of a third variable (*z*) driving both *x* and *y*, i.e., it does not say whether FC between these signals is direct or indirect. To solve this issue, partial mutual information (PMI) measures the amount of information shared by *x* and *y* while discounting the possibility that *z* drives both *x* and *y*. The partial mutual information between random variables *x,y* and *z* is defined by: *PMI(X,Y/Z) = H(X,Z) + H(Z,Y) - H(Z) - H(X,Z,Y)*, where H denotes the Shannon differential entropy (45). Thus, If *z* is independent of both *x* and *y*, *H(X,Y,Z)* equals 0 and PMI degenerates to MI (section 3.5.1) (Frenzel & Pompe, 2007).

### 3.5.3   *Transfer Entropy (TE)*

Assuming that two time series $x(t)$ and $y(t)$ can be approximated by Markov processes, a measure of causality that computes the deviation from the following generalized Markov condition was proposed (Schreiber 2000):

$$p(y_{t+1} \mid \mathbf{y_t^n}, \mathbf{x_t^m}) = p(y_{t+1} \mid \mathbf{y_t^n}) \qquad (48)$$

where $\mathbf{x_t^m} = (x_t, x_{t+1}..., x_{t-m+1})$ and $\mathbf{y_t^n} = (y_t, y_{t+1}..., y_{t-n+1})$ , being *m* and *n* orders (memory) of the Markov processes in *x* and *y*, respectively. The right hand side of (48) is the probability of obtaining a value of $y_{t+1}$ given its previous history *n* steps before, while the left hand side estimates this probability when both the histories of *x(t)* and *y(t)* are taken into account. Note that this is conceptually very similar to the idea of GC described in section 3.4. However, as commented later, transfer entropy does not assume a priori any kind of dependence (whether linear or nonlinear), and it is non-parametric. The price to pay in return for these advantages is that it is necessary to estimate probabilities from the data, which is normally not an easy task.

---

[5] To obtain an asymmetric (causal) estimation, delayed mutual information, (i.e. MI between one of the signals and a lagged version of another) has been proposed (Schreiber, 2000; Vastano & Swinney, 1988). This measure contains certain dynamical structure due to the time lag incorporated. Nevertheless, delayed mutual information has been pointed out to contain certain flaws such as problems due to a common history or shared information from a common input.

The equality above is fully satisfied when the transition probabilities (i.e., the dynamics) of *y* are independent of the past of *x*, that is, in the absence of causality from *x* to *y*. To measure the departure from this condition (and therefore the presence of causality), (Schreiber 2000) uses the Kullback-Leibler divergence between the two probability distributions at each side of (28) to define the transfer entropy from *x* to *y* as:

$$T_{X \to Y} = \sum_{y_{t+1}, \mathbf{y_t^n}, \mathbf{x_t^m}} p(y_{t+1} \,|\, \mathbf{y_t^n}, \mathbf{x_t^m}) \log \left( \frac{p(y_{t+1} \,|\, \mathbf{y_t^n}, \mathbf{x_t^m})}{p(y_{t+1} \,|\, \mathbf{y_t^n})} \right)$$

(49)

It measures the amount of directed information flow from *x* to *y*.

**DEFINITION:** Based on the definition above, the transfer entropy from time series $x_t$ to $y_t$ can be written as:

$$T_{X \to Y} = \sum_{y_{t+1}, \mathbf{y_t^{d_y}}, \mathbf{x_t^{d_x}}} p(y_{t+u} \,|\, \mathbf{y_t^{d_y}}, \mathbf{x_t^{d_x}}) \log \left( \frac{p(y_{t+u} \,|\, \mathbf{y_t^{d_y}}, \mathbf{x_t^{d_x}})}{p(y_{t+u} \,|\, \mathbf{y_t^{d_y}})} \right)$$

(50)

where *t* is a discrete valued time-index and *u* denotes the prediction time, a discrete valued time-interval. Besides, $\mathbf{y_t^{d_y}}$ and $\mathbf{x_t^{d_x}}$ are $d_x$- and $d_y$-dimensional delay vectors, as detailed below:

$$\mathbf{x_t^{d_x}} = \left( x(t), x(t-\tau), ..., x(t-(d_x-1)\tau) \right)$$
$$\mathbf{y_t^{d_y}} = \left( y(t), y(t-\tau), ..., y(t-(d_y-1)\tau) \right)$$

(51)

It is easy to see that (51) is equivalent to (22) if we take $d_x = d_y = d$, but here we explicitly take into account that the embedding dimensions (i.e., the memory of the Markov process in each signal) may be different for *x* and *y*.

HERMES computes transfer entropy between *x*, *y* and *w*, as a combination of entropies: *T(w,X,Y) = H(w,X) + H(X,Y) - H(X) - H(w,X,Y)*, where H is the Shannon entropy (45) and *w* is the future of *x*.

**RANGE:** $0 \leq TE_{X\to Y} < \infty$. (0): there is no causality between *x* and *y*, (>0): *x* is 'causing' *y*

**REMARKS**: Transfer entropy naturally incorporates directional and dynamical information, because it is inherently asymmetric and based on transition probabilities. Its main strength is that it does not assume any particular model for the interaction between the two systems of interest. Thus, the sensitivity of transfer entropy to correlations of all order becomes an advantage for exploratory analyses over GC or other model based approaches. This is particularly relevant when the detection of some unknown non-linear interaction is required (Vicente et al. 2011).

### 3.5.4   *Partial Transfer Entropy (PTE)*

As in the case of MI, it is possible to define a partialized version of TE, the so-called Partial transfer entropy (PTE), which measures the amount of directed information flow from $x$ to $y$ while discounting the possibility that $z$ drives both $x$ and $y$. The partial transfer entropy between the random variables $x$, $y$, $z$, and $w$ is defined as $PTE(w,X,Y|Z) = H(w,X,Z) + H(X,Z,Y) - H(X,Z) - H(w,X,Z,Y)$, where H is the differential entropy (45) and $w$ is the future of $x$. If $z$ is independent of both $x$ and $y$, then PTE degenerates to TE (section 3.5.3)

**REMARKS**: In the calculation of both PMI (section 3.5.2) and PTE, it is possible to take into account more than one additional (i.e., other than $z$) variable. Thus, for instance, PMI between $x$ and $y$ can be further partialized by removing the effect (if any) of a fourth (fifth, sixth,..) variable(s) $z_1$ ($z_2$, $z_3$...). This is accomplished simply by considering Z as a multidimensional random variable composed by several unidimensional random variables ( Z = $z_1$, ( $z_2$, $z_3$... ) ). However, in the current version of the toolbox, the calculation of PMI and PTE is possible for data containing up to ten variables, because the computational time necessary to calculate them in larger data sets is almost prohibitive. Besides, if the number of variables is high, it may better to use a recently derived method based on probabilistic graphic models (Runge et al., 2012a;b), which has been shown to outperform PTE in distinguishing direct from indirect connections in multivariate data sets.

### 3.5.5   *Parameters for the Information Theoretic measures*

1) **Embedding dimension (d) and Embedding delay (τ), for TE and PTE**: the same ranges, default values and remarks already mentioned in section 3.3.8 are applicable here.
2) **Mass of the nearest neighbour search (k):** level of bias and statistical error of the estimate. RANGE: [*d, 2d*]. DEFAULT: 4, according to Kraskov et al. (2004).

**REMARKS**: All the information theory based indexes described in this section (3.5), are computed using the code included in TIM 1.2.0 (http://www.tut.fi/tim), developed by Kalle Rutanen (Gomez-Herrero et al., 2010).

## 3.6 Summary table

Table 2 summarizes all the connectivity indexes included in HERMES along with a detailed description of their properties.

| Index/Characteristic | Non linear | Detect causal relations | Normalized | Discard indirect links |
|---|---|---|---|---|
| COR | | | ✓ | |
| XCOR | | | ✓ | |
| COH | | | ✓ | |
| PSI | | ✓ | | |
| PLV | ✓ | | ✓ | |
| PLI | ✓ | | ✓ | ✓ |
| WPLI | ✓ | | ✓ | ✓ |
| RHO | ✓ | | ✓ | |
| DPI | ✓ | ✓ | ✓ | |
| S | ✓ | | ✓ | |
| H | ✓ | | | |
| N | ✓ | | ✓ | |
| M | ✓ | | ✓ | |
| L | ✓ | | ✓ | |
| SL | ✓ | | ✓ | |
| GC | | ✓ | | |
| PDC | | ✓ | ✓ | ✓ |
| DTF | | ✓ | ✓ | |
| MI | ✓ | | | |
| TE | ✓ | ✓ | | |
| PMI | ✓ | | ✓ | ✓ |
| PTE | ✓ | ✓ | ✓ | ✓ |

**Table 2 Summary of the characteristics of the different connectivity indexes**

The list of indexes covered in the current version of HERMES, indicating also which characteristics has each index

# 4. STATISTICAL SIGNIFICANCE OF THE INDEXES

To evaluate the statistical significance of the indexes, HERMES includes the option of computing different surrogate data tests. The surrogate data method was introduced into practice two decades ago (Theiler, et al., 1992) and it is nowadays the most popular test for non-linearity in experimental data. It belongs to a more general type of statistical tests known as hypothesis tests (see, e.g., (Andrzejak et al., 2003) or Appendix A in (Pereda et al., 2005)).

Its usefulness in the field of connectivity analysis lies in the fact that sometimes the value of a FC / EC index is not due to the existence of statistical or causal relationship between the time series, but is the result of some feature of the individual signals (such as their complexity, their limited length of their non-stationarity, (Bhattacharya et al., 2003; Pereda et al., 2001; Quian Quiroga et al., 2000)). To test whether an index is actually measuring interdependence, multivariate surrogate data can be constructed, compatible with the null hypothesis that the signals are independent, and the significance of the corresponding indexes can be tested by comparing their values from the original data to those from these surrogate data.

## 4.1 Surrogate data for the nonlinear synchronization indexes

Surrogate data maintaining the amplitudes but destroying the phase relationship (if any) between the original signals are computed for the PS and GC indexes. The phases of the signals in the frequency domain are randomized by adding the same random quantity to the phases of each signal at each frequency. The procedure is carried out in the frequency domain, shuffling symmetrically the data, to obtain a real signal when transforming back to the time domain. More refined tests have been described in the literature to test, e.g., for PS. In particular, we mention the method of twin surrogates (Thiel et al., 2006), which is included in the Cross Recurrence Plot toolbox (Marwan et al., 2007) and the recently derived time reversed surrogates, which are very useful in combination with GC measures, to discard indirect connections in sensor space (Haufe et al., 2013). Additionally, the well-known TISEAN package (Hegger et al., 1999) includes also routines to generate different types of uni- and multivariate surrogate data.

## 4.2 Surrogate data for amplitude and phase

Surrogate data where any correlation in phase or amplitude has been removed (by randomly shuffling the time samples) are generated to estimate the statistical significance of the classical indexes.

### 4.2.1 Parameters for the surrogate data

For every index, there is the possibility of performing a surrogate data test, as described above, by selecting the option in their parameters panel. Once selected, you can indicate the number of surrogates to compute.

**Number of surrogates:** to have a *p*-value equal to *P* for the confidence of your test, you will need at least $1/P$ surrogates. A usual value for *P* is 0.05. In this case, by applying non-parametric rank statistics (the most conservative but often also the most realistic choice (Schreiber & Schmitz, 2000)), it will be necessary that only 1 out of the 20 values of the index for the surrogate data is lower than your original index value for the test to be significant. If you select *p*<0.01, then 100 surrogates are needed, and only 1 of the indexes for the surrogate data could be higher than the original, and so on. RANGE: [20,10000]. Please, bear in mind that the higher number of surrogates, the longer the computation will take. DEFAULT: 100, to get a *p*-value of 0.01.

## 4.3   How are the results from the surrogate data taken into account?

For a given connectivity index, the results for the *M* variables (whether electrodes, MEG sensors or sources) are stored in the interdependence matrix *A* for each subject. Thus, the element $a^k_{ij}$ is the interdependence index between variables *i* and *j* (*i,j=1,..,M*) for subject *k*. If the user decides not to perform the surrogate data test, these values are directly used in the statistical test described in the next section. If, however, he/she decides to make use of the multivariate surrogate data test, then an additional matrix *B* (a "mask") is produced for each subject, in the following way:

$$b^k_{ij} = \begin{cases} 0, & \textit{if } H_0 \text{ cannot be rejected} \\ 1, & \textit{if } H_0 \text{ can be rejected} \end{cases} \tag{52}$$

where $H_0$ is the null hypothesis of independence of the time series, which is rejected (or not) at the selected *p* level of statistical significance.

Then, a matrix $c^k_{ij} = a^k_{ij} \, b^k_{ij}$ is produced, and finally the values fed to the statistical test are $c^k_{ij}$. In other words, the effect of the surrogate data test is setting to zero all the non-significant original indexes. This can be rephrased by saying that the statistical test is always performed on the $c^k_{ij}$ values, and that, should the surrogate data test be not applied, $b^k_{ij} = 1; \forall \, i, j, k$.

## 5.   STATISTICAL TEST FOR MULTIPLE COMPARISONS

The statistical analysis of functional connectivity measures in EEG/MEG measurements is a challenging task, given the huge amount of information that must be subjected to statistical testing. If connectivity measures computed on recordings from two different subject groups are naively compared at the single sensor-pair level, the sheer largeness of the number of comparisons (which may have to be multiplied by the number of times windows and/or frequency bands taken into consideration) makes the finding of individually significant tests not only in conflict with the "p < 0.05" statement, but at times almost irrelevant, in the sense that the probability of not finding one single individual may rapidly approach zero as the number of simultaneous tests being performed increases. This is the well-known multiple comparisons problem, which has received much attention in the statistical literature (see, for instance, (Tukey, 1991) for a review and (Curran-Everett, 2000) for its importance in the context of biomedical statistical analyses and a review of some popular solutions).

HERMES includes the possibility of computing false discovery rate (section 5.1) and non-parametric cluster-based permutation test (section 5.2) between two different groups or conditions for a given index. Both tests can be selected in '*statistics*' panel (see Fig. 4).

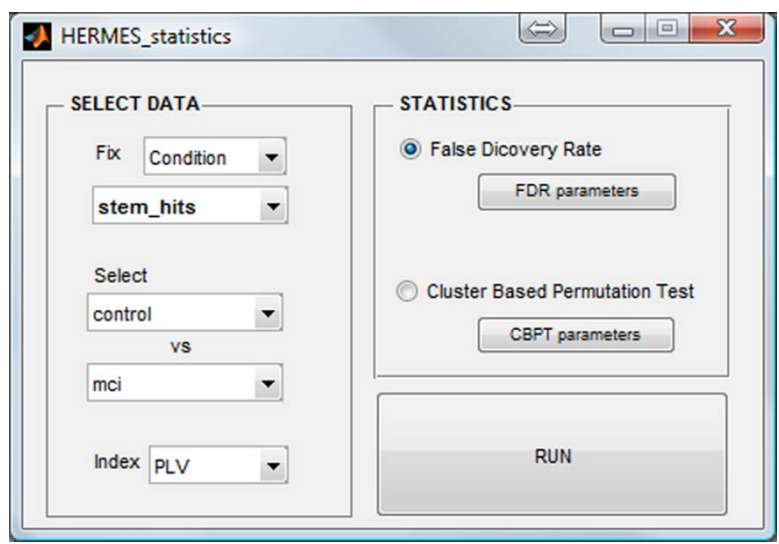

**Fig. 4 Statistics panel**

Statistics panel to select the condition to be compared (either between- or within- group), the index and the statistical test (with its corresponding parameters)

## 5.1 False Discovery Rate (FDR)

HERMES allows the computation of statistical tests between different groups and conditions, which take into account the abovementioned problem of multiple comparisons. One of the statistical methods used for multiple hypothesis testing is false discovery rate, which corrects for multiple comparisons (Benjamini & Yekutieli, 2001; Genovese et al., 2002) and could be applied to find channel pairs with significant differences between two groups of subjects. In a list of rejected hypotheses, FDR controls the expected proportion of incorrectly rejected null hypotheses (type I errors). It is a less conservative procedure for comparison than the Bonferroni correction, with greater power than familywise error rate control, at a cost of increasing the likelihood of obtaining type I errors.

Briefly, the FDR is the expected proportion of false positives among all significant hypotheses. For each channel pair, a between-group t-test (or its non-parametric equivalent, the Wilcoxon test) is calculated. From the resulting *p* values, a significance threshold was calculated with a corresponding *q* (typically *q = 0.2*, (Genovese et al., 2002) using the type I FDR implementation. The *q*-value is the expected fraction of false positives out of all positives and is the FDR analogue of the *p*-value of a typical hypothesis test. To understand how the method works, consider the problem of testing simultaneously *m* (null) hypotheses (Benjamini & Hochberg, 1995), of which $m_o$ are true. *R* is the number of hypotheses rejected, as summarized in Table 3.

|  | Actual values | Declared non-significant | Declared significant |
|---|---|---|---|
| **True null hypotheses** | $m_0$ | U | V |
| **Non-true null hypotheses** | $m - m_0$ | T | S |
| **Total** | m | m - R | R |

**Table 3 False discovery rate**

Explanation of the variables involved in the FDR method.

The specific *m* hypotheses are assumed to be known in advance, and *R* is an observable random variable, whereas U, V, S and T are unobservable random variables. The proportion of erroneously rejected null hypotheses is: Q = V/(V+S). Q is an unknown random variable, as we do not know V and S, even after experimentation and data analysis. We define the FDR $Q_e$ as the expectation of Q: $Q_e = E(Q) = E\{V/(V+S)\}$.

### 5.1.1   *Type I vs. Type II FDR*

Genovese et al., (2001) describe two methods for thresholding of statistical maps in functional neuroimaging, which easily adapts to the case of EC/FC analysis. In brief, for *V* voxels/sensor/sources being tested, a constant c(V) is defined for FDR type I (c(V) = 1) or type II (c(V) = $\sum_{i=1}^{N_c} 1/i$ ), where $N_c$ is the total number of performed comparisons[6]. Then, one proceeds as follows:

1.  First, select a desired FDR *q*-value between 0 and 1.This is the maximum FDR that the researcher is willing to tolerate on average.

---

[6] In the original paper by Genovese et al. (2001), Nc=V, as they study situations where only one comparison per voxel is performed to determine whether the voxel is significantly activated in a given group/situation. However, in the case of FC/EC indexes, Nc is the number of different indexes, which equals V(V-1)/2 and V(V-1) for symmetric and asymmetric indexes, respectively

2. Then, order the $p$ values from smallest to largest: $p(1) \leq p(2) \leq \ldots \leq p(N_c)$. Now, let $v(i)$ be the comparison corresponding to the value $p(i)$ and $r$ the largest $i$ for which $p(i) \leq \dfrac{iq}{c(V)N_C}$

3. Finally, declare the comparisons $v(1), \ldots , v(r)$ significant, or in other words, threshold the test statistics at the value $p(r)$ .

The choice of $c(V)$ depends on assumptions about the joint distribution of the $p$ values. The type II choice applies for *any* joint distribution of the $p$ values, whereas the type I choice ($c(V) = 1$) applies under slightly more restrictive assumptions: it holds when the $p$ values are independent and under a technical condition, called positive dependence, when the noise in the data is Gaussian with non-negative correlation across sensor/voxels (Benjamini & Yekutieli, 2001).

Since, the constant in type II is larger than that in type I, the corresponding cut-off for significance and number of comparisons declared significant is smaller in type II FDR.

### 5.1.2    Parameters for FDR test

1) **q:** minimum FDR at which the test may be called significant. RANGE: [0.01, 0.4]. DEFAULT: 0.2.
2) **Type**: '*Type I*' is more permissive, '*Type II*' is harder, more restrictive. DEFAULT: *Type II*.
3) **Method**: i) '*t-test*' for normal measures, ii) '*Wilcoxon*', its non-parametric equivalent. DEFAULT: *Wilcoxon*.

## 5.2    Cluster-Based Permutation Tests (CBPT)

The other possibility to perform statistical tests for multiple comparisons included in the current version of HERMES is the so-called nonparametric cluster-based permutation test (Maris & Oostenveld, 2007; Nichols & Holmes, 2002). This type of tests offers an intuitive and powerful nonparametric framework for the statistical analysis of connectivity measures (as well as signal amplitude measures, power spectra, etc.) that accounts for the inherent multiplicity involved in the statistical testing. The working assumptions of the variant of the method implemented in HERMES are relevant in a great majority of cases: if a statistical effect found in the comparison between groups or experimental conditions is to be considered significant, it should be 'larger' than the effects found when the measurements are randomly assigned to groups or conditions (which in principle can be accounted for by just the data structure, redundancy and multiplicity, and not by group membership or the performance of a task). 'Largeness' refers to a combination of the statistical strength of the effect and to its spatial largeness (and also its time duration or its frequency content, if the functional connectivity measure is sensitive to those dimensions as well), and its operational definition is linked with the concept of 'exceedance mass', as explained below.

The cluster-based permutation method included in HERMES works as follows: first, for each 'statistical unit' (the value of the functional connectivity measure for each sensor-pair or the combination of sensor-pair, a time window and/or a frequency band), a t-test is performed at the (uncorrected for multiplicity) p<0.05 significance level, which gives a naive first approximation to the study of statistical differences between groups or conditions. Units that pass this initial threshold are clustered together with those of their neighbours (adjacent units in space, but, whenever pertinent, also contiguous units in time or frequency) that also pass the test with consistent statistical differences (the associated t-values must have equal signs). Thus, for each of these clusters, the exceedance mass, defined as the sum of the t-values of all units belonging to a cluster (see, e.g., (Poline et al., 1997)), is computed, and the exceedance masses of the largest clusters are stored. The functional connectivity indexes from different data sets are then randomly assigned to two subsets of the same size as the original ones, thereby forming a division that is no longer faithful to the original real partition of the data sets into subject groups or experimental conditions, and the procedure is repeated hundreds or thousands of times. The exceedance mass of the largest cluster of each of these random realizations is computed, to be subsequently used as cluster-level statistic in the permutation test (see Ernst (2004) for a review of the permutation test rationale). For all the largest clusters in the original labelling of data sets into groups/conditions, a Monte-Carlo *p*-value is obtained according to the proportion of elements in the distribution of largest cluster exceedance masses from all random realizations exceeding the observed cluster-level test statistic. For example, if we generate 10000 of these random realizations and the exceedance mass of a given cluster is greater than that of the original partition for, say, 19 of such realizations, then the level of significance for this cluster is $p < 20/10000 = 0.002$.

A pair of nodes (channels) is considered to be adjacent to another pair of nodes (channels), whenever a node of the first pair is closer than the maximum distance to one of the nodes of the second pair, and the other node of the first pair is also closer than the maximum distance to the remaining node of the second pair. In any other case, the edges (the connectivity between those pairs of nodes) are not considered to be adjacent.

### 5.2.1    *Parameters for the cluster-based permutation test*

1) **Max distance coefficient:** CBPT spatially groups pairs of sensors, so a *maxdist* radius for considering two sensors as neighbours is needed. This *maxdist(i)* is computed as *coef\*min_dist(i),* where *min_dist(i)* is the minimum distance of *sensor(i)* to all the others. RANGE: [1, 3]. DEFAULT: 1.5.

2) **Alpha:** p-value for significant level. RANGE [0.001, 0.1]. DEFAULT: 0.05.

3) **Number of clusters:** RANGE: [1, 10]. DEFAULT: 10.

4) **Number of permutations:** RANGE: [20,10000]. Please, bear in mind that the higher number of permutations, the longer the computation will take. DEFAULT: 100, for a p-value of 0.01.

**REMARK:** For asymmetric (i.e., EC) indexes, HERMES works with the average of the two indexes (*x* to *y* and *y* to *x*).

# 6. DATA REPRESENTATION

Since the main purpose of the toolbox is to provide a user-friendly, easy way of calculating FC and EC indexes, we have not included in HERMES many options for data representation. Moreover, many popular freely available Matlab® toolboxes, such as *EEGLab* (Delorme & Makeig 2004; Delorme et al. 2011), *FieldTrip* (Oostenveld et al., 2011) , or *BrainStorm* (Tadel et al., 2011) already serve this purpose. Nevertheless, for the sake of completeness, we have also included in HERMES a Visualization menu, which can be accessed from the main panel, at the Visualization tab.

## 6.1 Signal visualization

Henceforth we show some screenshots of the output of the signal visualizations you can get with the Sensors 2D option. In this example, we use MEG recordings carried out with the 4D Neuroimaging MAGNES 2500 WH 148 MEG system. Data were acquired at a sampling rate of 250Hz, with on-line band-pass filter of 0.5–50 Hz. The task consisted of a modified version of the Sternberg's letter-probe task (Bajo et al., 2012). Data is only of hits. To have an equal number of epochs across participants, 35 epochs (0.9 s each one) were randomly chosen from each of the participants. We study 10 subjects: 5 from the control group and 5 from the mild cognitive impairment (MCI) group. We include these ten subjects at the 'Example data' section of the webpage (and its .her file with the computed measures to import it easily). Fig. 5 shows the raw data in each sensor location, whereas Fig. 6 shows the corresponding power spectra.

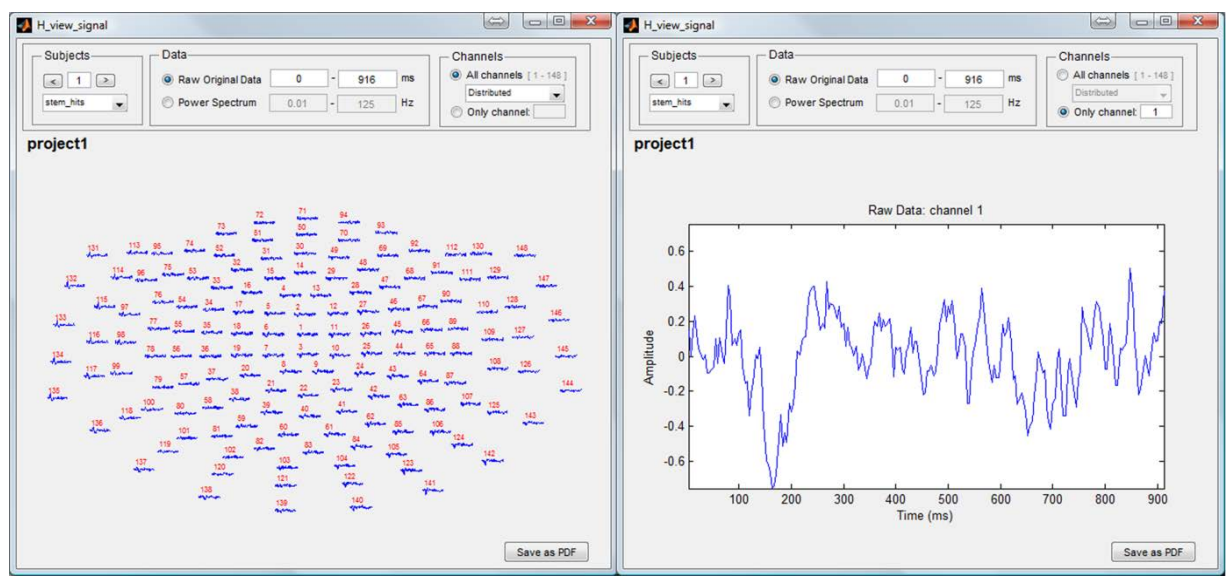

**Fig. 5 Raw data**

*Left:* Raw data of the whole set of sensors at each location. *Right:* Zooming into one single sensor (#1). Time range and channels' distribution can be adjusted. Subject's number and condition can be selected too

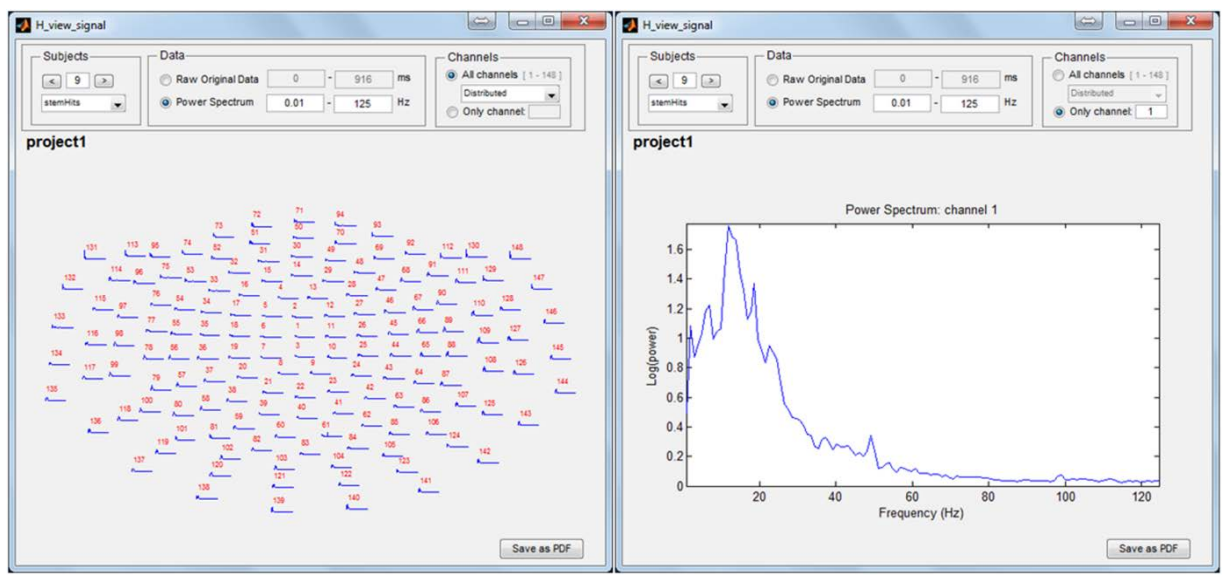

**Fig. 6 Power spectra**

*Left:* Power spectra of the whole set of sensors at each location. *Right:* Zooming into one single sensor (#1). As in Fig. 5, frequency range and channels' distribution can be adjusted. Subject's number and condition can be selected too

## 6.2 Connectivity visualization

Likewise, we also present some screenshots of the visualization of connectivity indexes in 2D. There are two main ways of visualization: Fig. 7 *left* presents the values of the PLV (section 3.2.1), a PS index of FC. In turn, Fig. 7 *right* presents the values of Granger Causality (section 3.4.1), an asymmetric linear index estimating EC between two sensors.

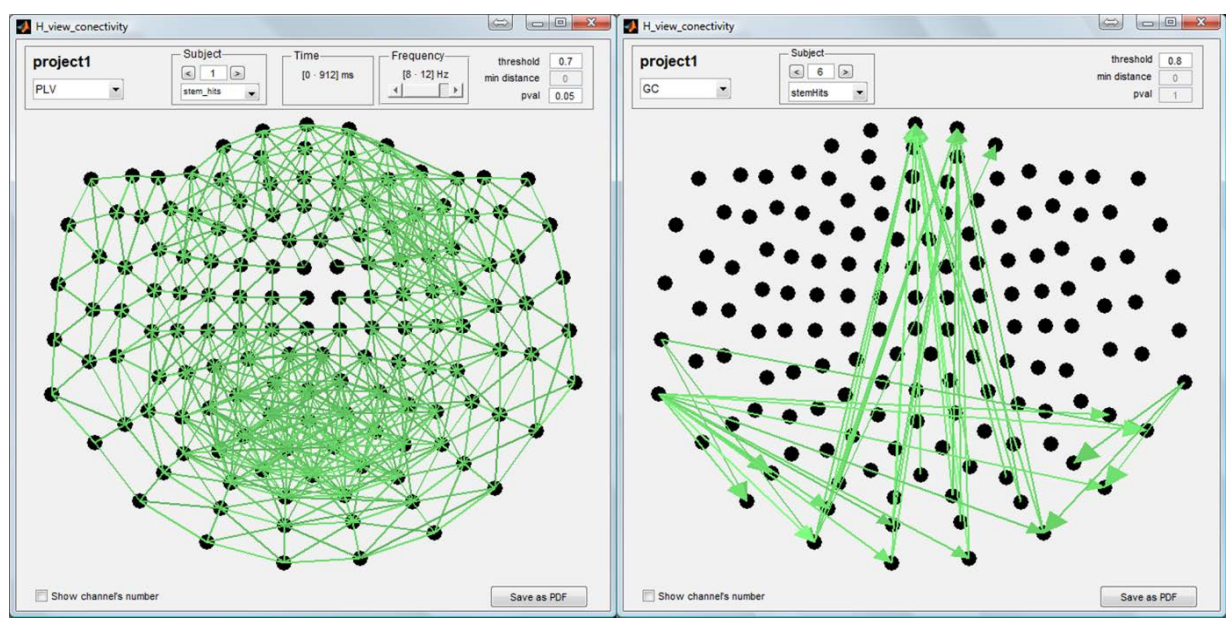

**Fig. 7 2D visualization of connectivity indexes for a 148 MEG sensor cap.**

*Left:* 2D visualization of the PLV, a PS symmetric index of FC. The segments between two sensors indicate that there exists FC, as assessed by this index, at the significance level considered (p<0.05, uncorrected for multiple comparisons). Different time and frequency ranges can be explored. Subject's number, group and condition can be also adjusted. *Right:* 2D visualization of GC asymmetric index. The arrows between two sensors indicate that there exists EC, as assessed by this index, at the significance level considered (p<0.05, uncorrected for multiple comparisons). The head of the arrow points to the "effect" sensor. Here too, different time and frequency ranges can be explored. Subject's number, group and condition can be also adjusted.

You can navigate in the time and frequency axis, and select the subject, the group, the type of experimental condition and the index you want to visualize. Moreover, you can also modify another three parameters:

- **Threshold:** it refers to the minimal value of the synchronization index from which all the other values are represented.
- **Minimum distance:** it refers to the minimum distance between two sensors (in 3D) from which links are represented (if no 3D coordinates are loaded this parameter is disabled).
- **p-value:** in the case you performed surrogate data tests for you indexes, this parameter is enabled, and reflects the degree of confidence you have in the index (as explained in section 4, STATISTICAL SIGNIFICANCE OF THE INDEXES), being the closer to 0, the more significant the values of the indexes.

You have the option of saving the images as a PDF file. All the images are appended to the same file, which is saved with the name: 'H_projectname.pdf' in the directory HERMES/Projects.

## 6.3   Statistics visualization

Statistically significant clusters can be visualized using the '*Statistics visualization*' panel. There, you can select the specific statistics results computed for a certain index with the chosen statistic method. Links where connectivity is stronger for the control group are represented in red, whereas if connectivity is stronger in the MCI group, they are represented in blue (Fig. 8).

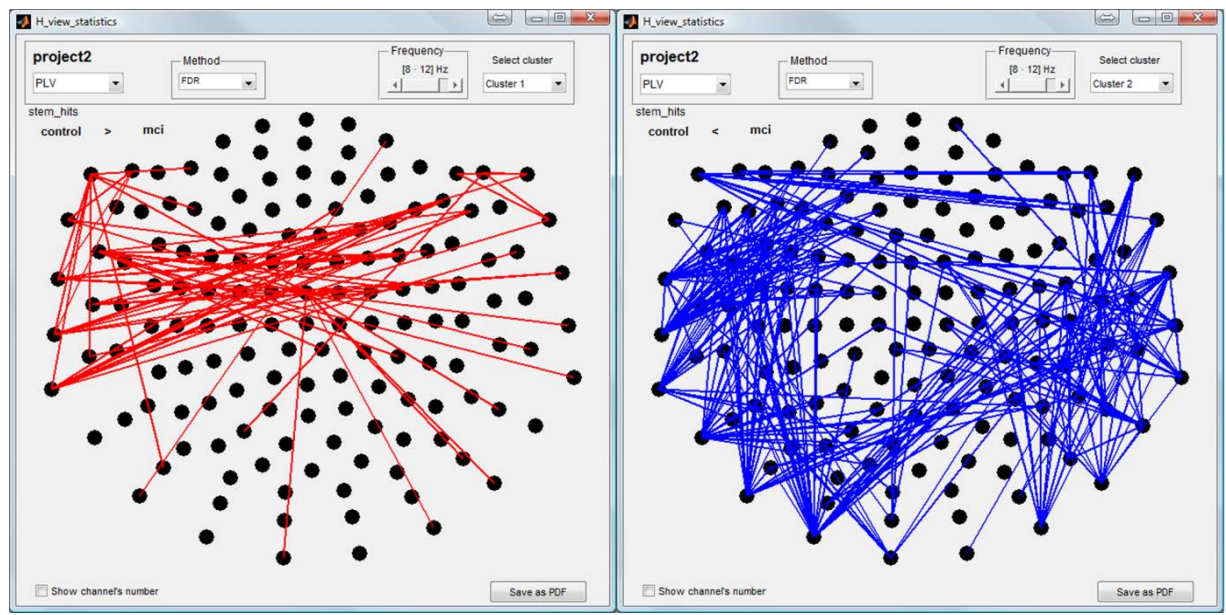

**Fig. 8 Statistics visualization**

2D visualization of the statistical differences for the PLV index between two groups of 5 subjects, as described in the text: healthy controls and mild cognitive impairment (MCI) subjects. A segment between two sensors indicates the existence of significant between-groups differences (type I FDR method, with p<0.05, q=0.3). In each panel, the group that presents a higher PLV value is indicated. *Left*: Pair of sensors presenting higher PLV values for the '*control*' group (*red segments*). *Right*: Same but for the '*mci*' group (*blue segments*)

# 7. DISCUSSION

In this work we have presented HERMES, a new Matlab® toolbox for the analysis of brain connectivity from neuroimage data. In contrast to existing toolboxes (Delorme & Makeig 2004; Zhou et al. 2009; Seth 2010; Tadel et al. 2011), HERMES encompasses both linear and nonlinear indexes of functional and effective connectivity, visualization tools and two different types of statistical tests to address the multiple comparisons problem. The main aim of the toolbox is to make the analysis of brain connectivity accessible to a wide community of researchers who are not familiar (and/or comfortable) with the mathematical and computational complexity of connectivity measures, that is why we have opted for producing a (hopefully) user-friendly GUI, from which all the calculations can be performed. Both spontaneous activity and event related (evoked) responses with many trial per subjects can be analysed, and the type of data that to be analysed are any kind of neuroimage data where several (at least $M=2$, but clearly $M>>2$ is the most common case) channels/sensors/sources/voxels are simultaneously recorded. This includes (but is not limited to) EEG (surface of intracranially recorded), MEG and fMRI data, with the only limitation of the minimal length of the time series necessary for the proper calculation of each selected index. Additionally, and fully aware of the current popularity of related toolboxes that include preprocessing methods yet have their own data structure (e.g., (Oostenveld et al., 2011), we allow HERMES to deal not only with raw data, but also with these special data structures, to ease the work of those researchers who want to make use of both toolboxes.

As commented in the INTRODUCTION, we feel that such a toolbox as HERMES was called for, as it includes many of the FC and EC indexes most popular in the literature while at the same time keeps them simple and easy to deal with, thanks to the GUI environment, without losing the necessary rigor. On making the effort of producing HERMES, we had in mind cases such as that of the TISEAN package (Hegger et al., 1999), which greatly helped to popularize the nonlinear analysis of time series by making publicly and freely available to the community a set of routines for the preprocessing and calculation of several nonlinear indexes from time series. Certainly, it would be very ambitious to expect that HERMES become as popular as TISEAN, the latter having being cited in no less than 630 papers as of May, 2013. But we do believe that our toolbox, even though intended for proprietary (yet extensively used) software such as Matlab® -instead of being simply Fortran or C++ code as in the case of TISEAN-, has the potential to be useful to a wide group of researchers of the neuroscientific community, who are willing to apply connectivity techniques to their data.

## 7.1   Future work

As the field of FC/EC analysis is a rapidly evolving one, for the toolbox not to become outdated it is necessary to keep on updating and upgrading it on a regular basis. Besides, advances in computational methods and computer hardware should be taken into account to improve the performance of HERMES and, specifically, speed up the computation of the indexes.

Regarding future upgrades/updates, we have already commented our intention of including in it all those new connectivity indexes that may be relevant (such as the recently published directed PLI (Stam & van Straaten, 2012a) cited in section 3.2.2 or the methodology cited in section 3.5.4 that outperforms PTE in certain situations (Runge et., 2012a)). Another two recently derived and potentially very useful FC indexes that we are planning to add to HERMES are the power envelope correlation between orthogonal signals (Hipp et al., 2012) commented in section 3.1.4 and the MIC (Reshef et al., 2011) briefly described in 3.5.1. Besides these bivariate FC/EC indexes, we think it will be worthwhile to include in the toolbox truly multivariate indexes such as those described in, e.g., (Allefeld & Bialonski, 2007; Allefeld, Muler, & Kurths, 2007; Bialonski & Lehnertz, 2006), which estimate the collective FC of $M$ signals as a whole by analysing the structure of the interdependence matrix $A$, whose elements are one of the bivariate FC indexes between every pair of signals[7].

As for the reduction in computational time, we intend, in the short term, to speed up the calculation specially of the GS and the information theory-based indexes, which are those ones having the greatest computational cost, by programming them in C/C++ and producing corresponding MEX files to use them within the Matlab® environment. In the longer term, we want to adapt the calculation algorithms to take advantage of multi-core hardware architectures and the Matlab® parallel computing toolbox. Both improvements will allow

---

[7] This is conceptually similar to the currently very popular approach of regarding A as the adjacency matrix of a complex network which is then used to calculate the main parameters of the network (Bullmore & Sporns, 2009; Stam & van Straaten, 2012b). However, such an approach will not be covered in HERMES, as there is already at least one comprehensive Matlab® toolbox for this purpose (Rubinov & Sporns, 2010)

the estimation of FC/EC patterns with HERMES for large number of channels/sensors and long time series in reasonable times.

# INFORMATION SHARING STATEMENT

HERMES (©UPM-ULL. 2012) is free software for academic use, distributed under the terms of the GNU General Public License v3. HERMES Toolbox and documentation are available freely and open-source at http://hermes.ctb.upm.es. Moreover, we have also made available for download in the website two sets of data, which we think may be useful. The first one contains the sample data from MEG recordings used in section 6. The second one consists of a single compressed file that includes several MAT files, each of them including two time series ($x$ variables) from two coupled dynamical systems (one Rössler system driving one Lorenz system, as depicted in (Quiroga et al., 2000; see eq. 11 and 12 herein)). The systems were integrated numerically for different values of the coupling parameter ranging from 0 (no coupling) to 1, as explained in the ASCII file that is included in the compressed file.

# ACKNOWLEDGEMENTS

We are very grateful to the generosity of many researchers who have made their code publicly available thereby contributing the initial seeds for HERMES. Among them, we are especially thankful to Dr. German Gómez-Herrero. We also acknowledge the role of Mr. Fernando Sanz, a former researcher of the Centre of Biomedical Technology, in the initial development of the toolbox.

We are also very grateful to the reviewers who perusal the original version of the manuscript and served as high quality beta testers of the toolbox, because they provided very useful feedback and comments that helped improving the quality of the work.

# ROLE OF THE FUNDING SOURCE

The authors acknowledge the financial support of the Spanish Ministry of Economy and Competitiveness through grants TEC2012-38453-CO4-01 and -03 and grant PSI2012-38375-C03-01 and the support of the Spanish Ministry of Science through grant PSI2009-14415-C03-01. Guiomar Niso has received the financial support of the Spanish Ministry of Education and Science through the FPU grant AP2008-02383. These funding sources have played no role in study design, the collection, analysis, and interpretation of data; in the writing of the report; and in the decision to submit the paper for publication.

# APPENDIX: HERMES DATA STRUCTURE

## A.1    Saved results

In HERMES, each created or imported project is stored in the folder *Results* in HERMES path. As commented in section 2, PROJECT CREATION, each project is stored in a folder, named after it by replacing the non-valid characters for the underscore character. Inside the project folder, several files and folders are stored at different levels. The directory tree or these files is detailed in Fig. 9.

According to this, the main project folder contains a mat-file named project, containing the main project structure; a file named *indexes.data*, containing the calculated indexes; and a series of folders named in the form *subject{n}*, with *{n}* correlative integer numbers in zero-padded three digits form.

Each *subject{n}* folder contains the time series of a subject, with data for each condition stored separately in a mat-file with the name *condition{n}.data*, being *{n}* correlative integers indicating the condition index as stored in the project structure. The data in these files is stored in a normalized form, with zero mean and unity standard deviation. This standardization is required by the synchronization indexes to perform an accurate estimation of connectivity.

Each one of the condition files is analyzed separately, allowing the comparison among different subjects and conditions. The results of this analysis are stored in the *indexes.data* file, in a Matlab® structure with one entry for each calculated index. The data stored in each entry includes both the configuration used to calculate the indexes and the obtained value, for each subject and condition, along with the (optional) significance level.

## A.2    Project structure

The project mat-file contains only the metadata of the project. This file does not contain any time series or results, to minimize memory requirements. As already mentioned, both the time series and the calculated indexes are stored in separated mat-files. The project structure is created as a combination of smaller structures. The metadata of the project is stored in different levels, as show in Fig. 9.

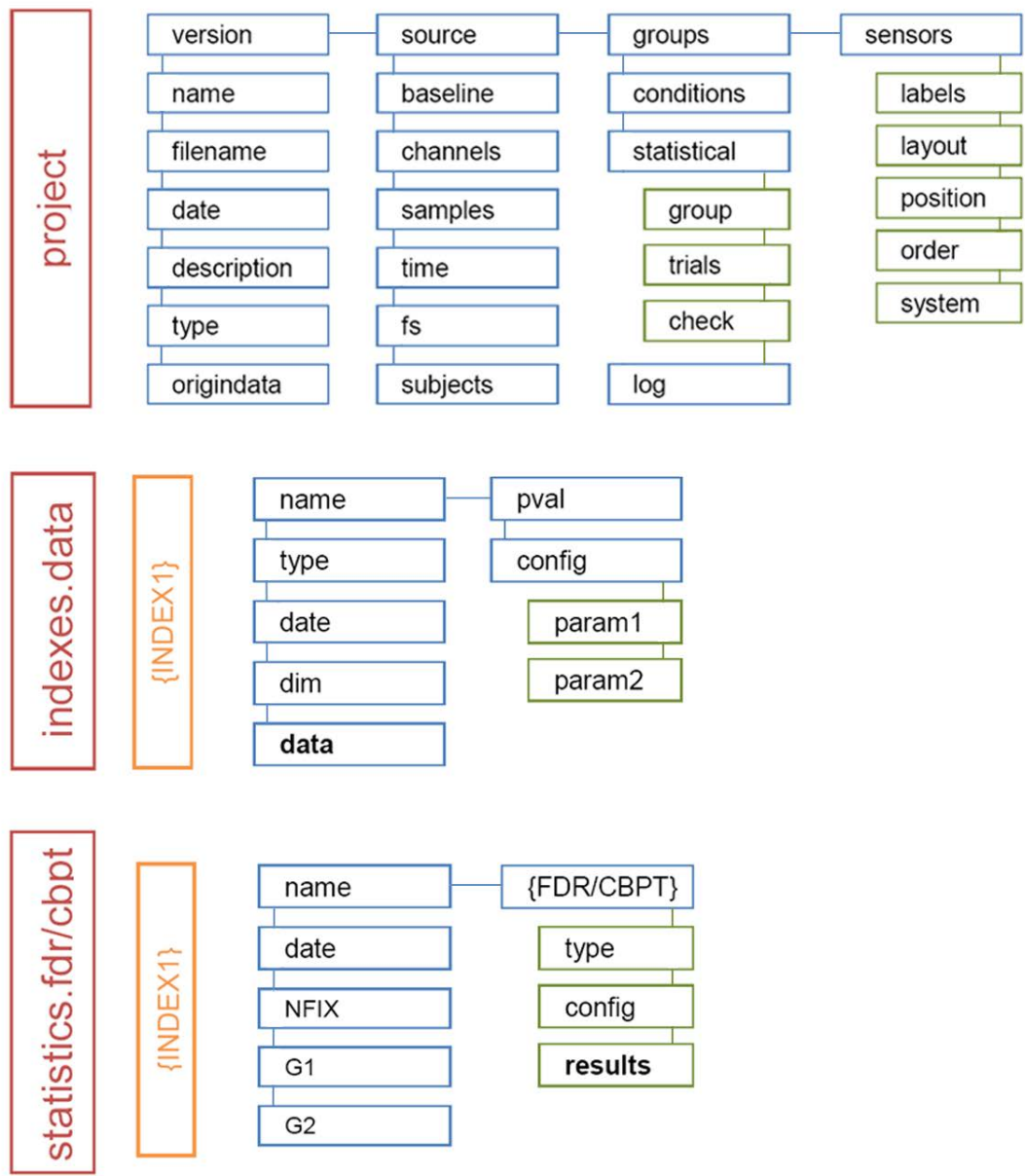

**Fig. 9 Project directory tree**

Structure of the two files in the *Projects* folder: *project* and *indexes.data*

**project** mat-file: All the metadata information of the project is stored here.

- *project.version*: String defining the version of HERMES where the project was created.
- *project.name*: Name of the project, assigned by the user.
- *project.filename*: Filename-friendly version of the project name, obtained by stripping out forbidden characters and replacing them by underscores.
- *project.date*: Date and time of the creation of the project, in Matlab® clock format.
- *project.description*: Description of the project provided for the user during the creation.

- *project.type*: Type of the project: 'continuous' registration or data 'with trials'.

- *project.origindata*: Acquisition metadata obtained from the original files.

- *project.source*: Origin from where the data were obtained.

- *project.baseline*: Baseline duration (in milliseconds) in data with trials. In continuous data this value is always 0.

- *project.channels*: Number of channels.

- *project.samples*: Number of samples of the epoch.

- *project.time*: Vector of the epoch times (in ms).

- *project.fs*: Sampling rate (in Hertzs).

- *project.subjects*: Cell array containing the name of each subject present in the project.

- *project.groups*: Cell array containing the name of each group present in the project.

- *project.conditions*: Cell array containing the name of each experimental condition present in the project.

- *project.statistical*: Multidimensional structure with as many elements as subjects in the project. The fields of the structure are:

  o *project.statistical.group*: The group ID of the subject as listed in the *project.groups* cell array.

  o *project.statistical.trials*: Array containing the number of trials/epochs for each condition. Each element in the array corresponds with one condition, as listed in the *project.conditions* cell array.

  o *project.statistical.check*: Checksums of each file for this subject. This value is used to check the integrity of the mat-files.

- *project.sensors*:

  o *project.sensors.label*: Labels indicating the name of the channels.

  o *project.sensors.position*: Tri-dimensional position of each channel, if known.

  o *project.sensors.layout*: Bi-dimensional position (and size) of the plots to draw the 2D layout and the connectivity plots.

  o *project.sensors.order*: List of channels selected from the mat-files and its order.

  o *project.sensors.system*: Name of the system layout selected in the creation of the project.

- *project.logs:* Multidimensional structure containing the metadata of the session logs created in the execution of HERMES for this project. The fields of the structure are:

  o *project.logs.filename*: String containing the name of the file where the session log is stored.

  o *project.logs.date*: Array containing the creation date and time of the session log.

  o *project.logs.description*: String containing a description of the log content.

**indexes.data** mat-file: This file contains the description and parameters of the indexes calculated for the dataset. Each index is stored in a separated variable, named after its abbreviation (e.g. COR for Correlation index or GC for Granger Causality index). These variables contain a structure *{index}*, with the fields:

**{index}**

- *{index}.name*: The complete name of the index and its abbreviation.
- *{index}.type*: The family of the index.
- *{index}.date*: Date and time when the index was calculated, in Matlab® clock format.
- *{index}.dimensions*: Cell array containing the label of each dimension in the calculated matrix, along with the value of each step.
- *{index}.config*: Structure containing the configuration used to calculate the index.
- *{index}.data*: Cell array containing the computed results of the index for each subject and condition.
- *{index}.pval*: Cell array containing the estimated p-values matrix for the values of the index for each subject and condition.

**statistics.fdr** and **statistics.cbpt** mat-file: These files contain the description and parameters of the statistical tests calculated for a certain index of the dataset. Each index is stored in a separated variable, named after its abbreviation (e.g. COR for Correlation index or GC for Granger Causality index). These variables contain a structure *{index}*, with the following fields:

**{index}**

- *{index}.name*: The complete name of the statistical method used and its abbreviation.
- *{index}.date*: Date and time when the index was calculated, in Matlab® clock format.
- *{index}.NFIX:* Fix *'Groups'* or *'Conditions'*.
- *{index}.FIX:* Name of the group or condition fixed.
- *{index}.G1:* Name of one of the groups or conditions being compared.
- *{index}.G2:* Name of the other group or condition being compared.
- *{index}.{FDR or CBPT}:* Structure which contains the results obtained and the parameters used:
  - *type*: *'False discovery rate'* or *'Non parametric permutation test'*.
  - *config*: Structure containing the configuration used to calculate the statistics.
  - *results*: Results of the statistics for the selected groups.

The fields in this structure may vary along the versions of HERMES. Thus, a project created by a certain version cannot be readable by another one by directly copy the project folder inside the Projects directory. We encourage users to interchange projects between different runs of HERMES by using the export/import tools described in Section 2.3. The structure fields for each version of the toolbox will always be available in the Documents section of the HERMES website.